\documentclass{aastex}
\usepackage{emulateapj5}

\usepackage{natbib}

\shortauthors{Chiboucas et al.}
\begin{document}
%\pdfgraphics

\title{Keck/LRIS Spectroscopic Confirmation of Coma Cluster Dwarf Galaxy Membership
Assignments}

\author{Kristin Chiboucas\altaffilmark{1,2},  R. Brent Tully\altaffilmark{2}, Ronald O. Marzke\altaffilmark{3}, Neil Trentham\altaffilmark{4}, Henry C. Ferguson\altaffilmark{5}, Derek Hammer\altaffilmark{6}, David Carter\altaffilmark{7}, and Habib Khosroshahi\altaffilmark{8}}
\email{kchibouc@gemini.edu, tully@ifa.hawaii.edu, marzke@sfsu.edu, trentham@ast.cam.ac.uk, ferguson@stsci.edu, hammerd@pha.jhu.edu, dxc@astro.livjm.ac.uk, habib@ipm.ir}

\altaffiltext{1}{Gemini Observatory, 670 N. A'ohoku Pl, Hilo, HI 96720}
\altaffiltext{2}{Institute for Astronomy, University of Hawaii, 2680 Woodlawn Dr., Honolulu, HI 96821}
\altaffiltext{3}{Department of Physics and Astronomy, San Francisco State University, San Francisco, CA 94132-4163}
\altaffiltext{4}{Institute of Astronomy, Madingley Road, Cambridge CB3 0HA, UK}
\altaffiltext{5}{Space Telescope Science Institute, 3700 San Martin Drive, Baltimore, MD 21228, USA}
\altaffiltext{6}{Department of Physics and Astronomy, Johns Hopkins University, 3400 North Charles Street, Baltimore, MD 21218}
\altaffiltext{7}{Astrophysics Research Institute, Liverpool John Moores University, Twelve Quays House, Egerton Wharf, Birkenhead CH411LD, UK}
\altaffiltext{8}{School of Astronomy, Institute for Research in Fundamental Sciences (IPM), PO Box 19395-5531, Tehran, Iran}

\begin{abstract}

Keck/LRIS multi-object spectroscopy has been carried out on 140 of some of the lowest and
highest surface brightness faint ($19 < R < 22$) dwarf galaxy candidates
in the core region of the Coma Cluster.
These spectra are used to measure redshifts and establish
membership for these faint dwarf populations.
The primary goal of the low surface brightness sample is to test our ability
to use morphological and surface brightness criteria to distinguish between Coma Cluster
members and background galaxies using high resolution HST/ACS images.  
Candidates were rated as expected members, uncertain, or expected background.
From 93 spectra, 51 dwarf galaxy members and 20 background 
galaxies are identified.  Our morphological membership estimation success rate is
$\sim100$\% for objects expected to be members and better than $\sim90$\% for galaxies expected to
be in the background.   We confirm that low surface brightness is a very good indicator of cluster
membership.  High surface brightness galaxies are almost always background with confusion arising
only from the cases of the rare compact elliptical galaxies.  The more problematic cases 
occur at intermediate surface brightness.  Many of these galaxies are given uncertain 
membership ratings, and these were found to be members about half of the time.
Including color information will improve membership determination but 
will fail for some of the
same objects that are already mis-identified when using only surface brightness and
morphology criteria. Compact elliptical galaxies 
with $B-V$ colors $\sim 0.2$ magnitudes
redward of the red sequence in particular require spectroscopic follow-up.
In a sample of 47 high surface brightness, 
ultra compact dwarf candidates, 19 objects have redshifts which place them
in the Coma Cluster, while another 6 have questionable redshift measurements
but may also prove to be members.   
Redshift measurements are presented and the use of indirect means for establishing 
cluster membership is discussed.

\end{abstract}


%----------------------------------------------
\section{Introduction}\label{intro}
%----------------------------------------------

In order to study the dwarf galaxy population of clusters,
cluster membership must first be unambiguously determined. 
Only in the Local and other very nearby groups, such as M81,
can membership be determined directly for all potential members.
Few rich galaxy clusters are close enough for a complete census.  
Historically, astronomers have assigned membership probability 
based on indirect means such as colors or morphology, or have
resorted to statistical methods for estimating membership.  The
latter method estimates the number of galaxies belonging to a cluster  
by correcting the total galaxy counts in a cluster region for the 
expected background population based on galaxy counts in 
nearby or blank regions of the sky.  
This statistical method may be the least secure as it assumes 
nearby regions on the sky to have the same background structure as the 
cluster area, yet it is known that structure is not homogeneous 
and that counts can vary by $\sim20-25$\% on scales of $\sim0.5$ deg 
\citep{hjddgo00,lc95}.  
The situation may be worse around rich galaxy clusters which
are found at the intersections of filaments where the true background population may
be significantly larger than in nearby less dense regions of sky.
The statistical method therefore not only introduces large uncertainties
due to field-to-field variance, but may suffer from significant systematic
bias as well.  
It is expected that the use of colors
and morphology will do a better job, but this needs to be tested
through direct means.  It is prohibitively time consuming to
obtain sufficiently high S/N spectra to measure redshifts 
for every galaxy in a cluster region down to the regime of the
abundant faint dwarf galaxy population,
and impossible with our current technology to measure redshifts
of the faintest and lowest surface brightness objects. We can,
however, observe a subsample of spectra to test memberships
determined from indirect methods such as
surface brightness, morphology, and color criteria.

Using morphological criteria to establish membership involves weeding out objects
with structure resembling that of background late type galaxies,
and using surface brightness, scale length, and morphological class to constrain
the remainder.  It is expected that most cluster galaxies are quiescent early-type
galaxies that have relatively lower surface brightness as compared to 
background objects. In rich galaxy clusters, the dwarf ellipticals form the dominant
population at low luminosities.  These galaxies have smooth low surface brightness 
profiles and relatively large sizes compared to background galaxies with similar apparent
magnitudes \citep[e.g.][]{cb87, ibm88, fs88}.
Thus, at faint magnitudes, it is expected that cluster dwarf galaxies can 
be distinguished from background galaxies by such properties as surface brightness 
and size.  However, there is always the possibility that
these criteria could fail if, for example, there existed a large population
of very compact objects (e.g. compact dEs and UCDs) which reside within
the cluster but are indistinguishable, morphologically, from background
spheroidals or foreground stars, or if there existed a large population of
giant low surface brightness galaxies in the cluster background.  
Such possibilities can be tested with spectroscopic samples.

This work is part of the HST/ACS Coma Cluster Treasury Survey \citep{carter}, 
a two-passband imaging survey originally designed to cover 740 arcmin$^{2}$ in 
the Coma Cluster to a depth of $I_{C} \sim 26.6$ for point
sources and $\mu_{e,I} < 26.0$ mag arcsec$^{-2}$ for extended sources. 
Over 200 arcmin$^{2}$ have been completed. 
The Coma Cluster was chosen because it is the densest, richest
cluster of galaxies at high galactic latitude within 100 Mpc and
will provide the local and current analog of rich clusters seen at 
higher redshift.  In addition, much progress has already been made towards 
understanding the cluster dynamics and galaxy properties of this rich 
cluster.
The HST/ACS survey provides a database of high spatial resolution 
images with the overall goals to study the structure, stellar populations,
and morphologies of, and environmental effects on, the cluster dwarf galaxy
population to a regime fainter than could be explored before.
Specifically,
we are using the ACS data to study e.g. the faint end of the cluster 
galaxy luminosity function (Trentham et al. in prep), 
structural parameters of dwarf galaxies (Hoyos et al. in prep), 
stellar populations \citep{smith08}, 
the nature of compact and ultra compact dwarf galaxies \citep[Chiboucas et al. in prep]{compact},
and globular cluster systems (Puzia et al. in prep, \citet{peng09}).  
This study will moreover provide the local rich cluster benchmark for 
various scaling relations.

The high quality ACS data are used to study structural, photometric,
and morphological properties of the cluster members.  However, 
cluster members must first be identified.  Several members of the team have therefore
made use of morphological, surface brightness, and size criteria to assign
a membership probability to each object in our ACS images.
As a follow-up to this ACS imaging survey, we are therefore establishing 
membership for faint dwarf galaxies
through spectroscopy for a subset of objects suspected of being cluster
members.  A separate MMT/Hectospec campaign has measured redshifts
for over 8700 galaxies down to $R \sim 20$, where it is $\sim 80$\% complete
for the ACS surveyed region. 
We are using Keck/LRIS to obtain spectra of even fainter galaxies
($19 < R < 24, -16 < M_R < -11$), targeting 
both the low and high surface brightness segments of the dwarf population. 

Galaxies at the faint end of the luminosity function tend to be low
surface brightness (LSB) dwarf ellipticals \citep{ttv01,fb94}.  To test our membership
assignments which follow from that assumption, we have observed samples
of both low surface brightness faint galaxies expected to be members and
higher surface brightness galaxies expected to lie in the background.  
We target the central core of the Coma Cluster where we have obtained
high resolution HST/ACS imaging and have completed morphological estimates for 
membership. 

At the very high surface brightness end of the spectrum lie the ultra
compact dwarfs (UCDs).  First discovered in spectroscopic studies of
the Fornax cluster, including an all-object spectroscopic
survey by \citet{djgp00}, significant UCD populations
have now been established to exist in the Fornax, Virgo, and Centaurus clusters
\citep{hil99,drink04,firth08,jones06,has05,mieske07,mieske09} while candidate UCDs
have been identified in the Hydra, A1689, AS0740, and Coma Clusters and the
NGC 1023 group \citep{wehner07,mieske04,blake08,ad09,mieske06}.  

UCDs have magnitudes in the range $-13.5 < M_V < -10.5$ and are
extremely compact, with physical sizes ranging from $7 < r_e < 100$ pc 
\citep{mieske07,jones06,mieskeb08}.  In the Coma Cluster, this range corresponds
to angular sizes of $0.01 - 0.2$ arcsec.  Coma UCDs are therefore 
unresolved in typical ground-based imaging, and only the largest can be 
resolved with HST/ACS. With high surface brightness and low luminosities, they are 
well distinguished from the magnitude - surface 
brightness relation of "typical" dwarf galaxies. UCDs also tend to be redder 
than the dE red sequence by $\sim0.2$ magnitudes.
This property is likely 
driven by high metallicities \citep{mieske06b,chil08}.  
The origin of these enigmatic objects is not clear. They 
may be the visible nuclei of extremely low surface brightness dwarf galaxies
or they may constitute a population of remnant nuclei formed from "threshing" of 
nucleated dwarf ellipticals via tidal encounters with giant galaxies \citep{pdg01,bekki03,has05}. 
Alternatively, they  may make up a bright tail of the globular 
cluster luminosity function or they may have formed as mergers of young massive star 
clusters themselves formed during galaxy mergers 
\citep{djgp00,fk02,chil08}.
The population size, spatial distribution, and intrinsic properties of this
object class, should UCDs exist within the Coma Cluster,
will provide critical new clues for understanding the formation 
mechanisms of these unusual
objects, and expand our understanding of the full dwarf galaxy population.

In Section \ref{obs} we detail the target selection, observations, and data
reduction.  We provide results of the spectroscopic study in  Section \ref{results}
including the success rates for using indirect means to estimate cluster membership.
In Section \ref{disc} we discuss these results and in Section \ref{conc} we provide
a summary and conclusions.
Throughout this paper, we assume a distance to the Coma Cluster of 100 Mpc
and a distance modulus $\mu = 35$ \citep[see][Table 1]{carter}.

%--------------------------------------------------------------
\section{Observations and Data Reduction}\label{obs}
%--------------------------------------------------------------

%--------------------------------------------------------------
\subsection{Target Selection}\label{targs}
%--------------------------------------------------------------

Our survey of faint dwarf galaxies in the Coma Cluster
was divided into the two surface brightness samples each with 
a slightly different science objective.
For the LSB sample the
goal is to use redshift measurements to test our ability to correctly determine
membership for Coma Cluster galaxies through indirect means.
This test provides statistical probabilities 
for dwarf galaxy cluster membership, which are needed to 
constrain the poorly known faint-end slope of the LF and to enable
dwarf population studies at the very faint regime.  The goal of the high
surface brightness sample is to investigate whether the Coma Cluster hosts a 
population of UCDs.

Prior to our initial observing run, redshifts existed for all galaxies in the
Coma region with $R < 17.5$ and for almost all expected members to
$R < 18.5$.  An ongoing MMT/Hectospec campaign has been targeting all galaxies
with $R < 20.0$.  Our plan was to take spectroscopic observations $2-3$ 
magnitudes deeper.
Our low surface brightness targets were therefore chosen within the magnitude 
range $19 < R < 22.5$, and with central surface brightnesses $\mu_o < 24$.  
$R-$band photometry was taken from the catalog of \citet{adami}.  At high
surface brightness, we targeted objects to $R < 24$. 

%--------------------------------------------------------------
\subsubsection{Morphological Membership Assessment}\label{LSBtargs}
%--------------------------------------------------------------

Indirect means are required to establish membership at faint magnitudes where
galaxy counts become too large for comprehensive spectroscopic coverage.
We therefore use known and expected morphological and structural properties of 
the cluster population in order to distinguish between background and 
cluster galaxies in our high resolution ACS imaging.  
This method has been used previously
for studies of other nearby galaxy groups and clusters.  One of the historically 
more prominent uses was the assembly of the Virgo Cluster Catalog using
primarily morphological means \citep{bst85}. Morphological classification
of galaxies in this region was performed by \citet{sb84} who define the classification
scheme and supply a reference
set of examples to serve as templates. Cluster membership was then determined primarily based
on the morphological classification for each galaxy.
More recently, this method has been used to assign membership to nearby group and
cluster galaxies in order to construct the galaxy luminosity functions \citep{ttv01,tt02,mtt05}.
These studies used a very similar method as this work to assign membership
to groups $\sim4$ times closer with $\sim4$ times lower spatial resolution from ground-based 
imaging.

The majority of faint galaxies imaged within a field of even a rich galaxy cluster
will be background galaxies.  Faint cluster members are the dwarf galaxies, and within a rich
cluster, the great majority of these will be of the early type dE class.
Both dwarf elliptical and dwarf irregular
galaxies follow a well defined magnitude - surface brightness relation 
where surface brightness decreases with
magnitude such that the physical radius of the galaxy remains nearly constant
\citep[e.g.][]{cb87,bc91,b94, mieske04b,bos08}.   
Thus, at a given apparent magnitude, dwarf cluster members are expected to be of lower 
surface brightness and to have larger sizes than the majority of background galaxies.  
The dwarf members are also expected to have roughly similar sizes.
Furthermore, late type field background galaxies can be distinguished based on morphological
and structural properties.  The presence of features such as spiral arms, dust lanes, 
evidence for tidal disruption, flat edge-on disks, 
central bulges, major bars, etc. are all good indicators for late type morphology more likely
to be associated with intrinsically high luminosity background galaxies than
intrinsically low luminosity dwarf members.

All ACS images from our survey were visually inspected by two of the authors
(NT and HF) who closely followed the \citet{sb84} prescription for galaxy classification.
Galaxies in the Coma Cluster fields were then placed into 4 membership categories based on
the following considerations. 
Objects at $R > 19$ that are compact and/or have spiral or other 
obvious structure indicative of intrinsically luminous galaxies were assumed 
to lie in the background.  From the magnitude - surface brightness relation,
faint members are expected to have
low surface brightness and relatively large sizes in comparison with the
abundant background galaxies of similar apparent magnitude.  These mostly featureless
diffuse LSB galaxies are assigned to the dE class and are assumed to be members. 
LSB galaxies with small central nuclei are morphologically assigned to
be of the dE,N class and can also be identified as members with relative 
certainty.  Besides the obvious members and background galaxies there will be some
cases where the membership cannot be unambiguously determined based on morphological
grounds alone.
The following probability ranking scheme, described further in Trentham et al. (in prep),
was then used to assign a membership likelihood to each galaxy: 

\begin{description}
\item [1:] Expected member
\item [2:] Probable member
\item [3:] Possible member  
\item [4:] Expected background
\end{description}

\begin{figure}[t]
\begin{centering}
\includegraphics[angle=0,totalheight=3.0in]{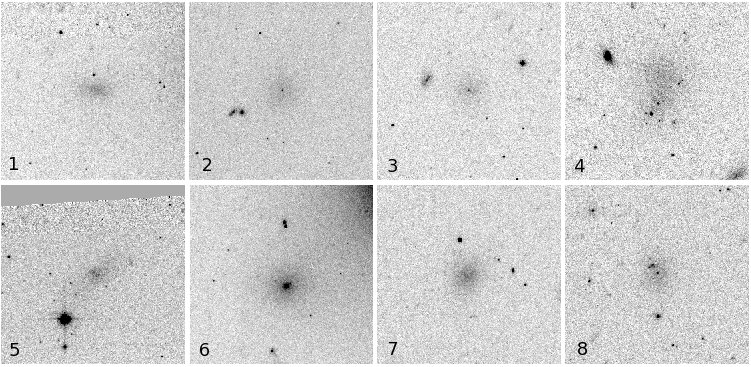}
%\plotone{fig1}
\caption[]{Examples of galaxies assigned membership probabilities 1-2.
Images are 18 arcsec across.
The top row displays 4 examples of galaxies considered definite members (ranking 1)
due to the smooth low surface brightness appearance of the dE and dE,Ns.  
In the bottom row are 4 galaxies considered to be very likely members (ranking 2).
Although very probable dE and dE,N cluster members, these have a slightly patchier
appearance.   
\label{class12}}
\end{centering}
\end{figure}

\begin{figure}[t]
\begin{centering}
%\plotone{fig2}
\includegraphics[angle=0,totalheight=3.0in]{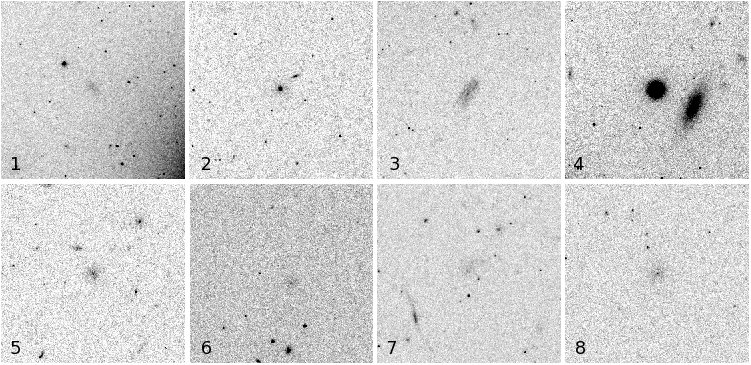}
\caption[]{Examples of galaxies assigned membership probability 3. Images are
18 arcsec across.
This is the most ambiguous rating, and includes galaxies that are considered possible 
member or background.  In each case it is apparent why the
galaxy is not clearly a background galaxy or a cluster member.  In panel (1), the object
has low surface brightness, but is small and elongated, panels (2) and (3) show possible
distant spirals, and panel (4) is a compact high surface brightness object, either
a more distant elliptical or a compact dwarf galaxy.  Galaxies such as this one
require spectroscopic confirmation of membership.  In the bottom row, all are 
similar to dE and dE,N types, but are smaller than other class 1 objects
and have a patchier appearance.
\label{class3}}
\end{centering}
\end{figure}

\begin{figure}[t]
\begin{centering}
\includegraphics[angle=0,totalheight=3.0in]{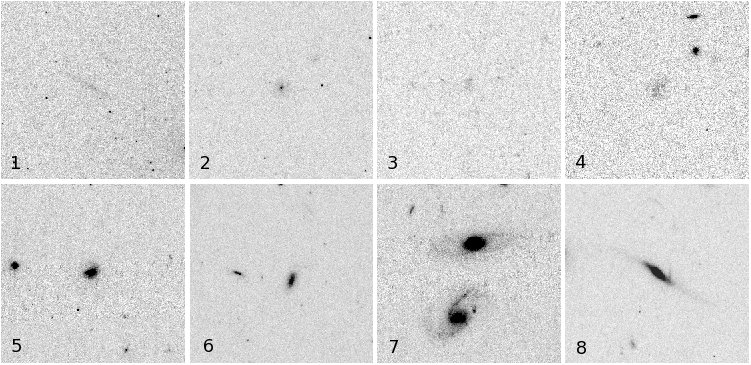}
%\plotone{fig3}
\caption[]{Examples of galaxies assigned membership probability 4.
Images are 18 arcsec across.
These galaxies are expected to lie in the background, due to features
that suggest late type morphologies and intrinsically high luminosities.
The top row displays
4 cases of low surface brightness galaxies expected to be late types at
large distance.  Identifying features include highly elliptical edge-on disks,
spiral structure, asymmetric profiles, and lumpy appearance.  In the bottom
row, these high surface brightness objects are believed to be 
late type field galaxies, most exhibiting clear spiral and tidal structures. 
\label{class4}}
\end{centering}
\end{figure}

Examples from each class are shown in Figures \ref{class12} - \ref{class4}.
We cannot provide a specific recipe for assigning memberships. This rating scheme is subjective
and therefore not perfectly reproducible.  However, 
it is expected that a human can be trained through experience looking at many images 
in different contexts to morphologically classify galaxies and reliably 
pick out members from qualitative determinations better than any devised algorithm. Both
authors here have significant experience with this technique.   Given previous
success using this method to determine membership for nearby group and cluster members,
it is expected that this method will have good success in assigning membership
for the more distant Coma Cluster galaxies (100 Mpc) using high resolution HST/ACS images.  
We test this hypothesis in this paper.

To test the reliability of these membership assignments, we chose samples
from each of the 4 membership classes for follow-up spectroscopy. 
We weighted all possible targets based on membership probability, 
giving higher weights to those galaxies considered likely
and probable members (ranking 1-2), a lower weighting for those considered possible
members (3), and the lowest weighting for those considered background (4). These
weights were used strictly in the LRIS mask making software, Autoslit3, and
were implemented primarily to obtain roughly equal numbers of each category in the
final masks, since total numbers of objects increased inversely with
membership likelihood.  In the end, $\sim20-25$ objects were chosen from each 
of the membership classes.
In order to ensure that the majority of spectra were of high enough
S/N to measure redshifts, we also gave slightly greater weight to those objects
with higher surface brightness. While this is a selection
bias, the main goal was to obtain high enough S/N for as many
targets as possible from which to extract secure redshift measurements.  
The sample for which we successfully measured redshifts has a large enough range
to investigate any possible bias or trend as a function of surface brightness.
We discuss this further in Section \ref{disc}. 

%--------------------------------------------------------------
\subsubsection{UCD Candidate Selection}\label{ucdtargs}
%--------------------------------------------------------------

The second goal of our spectroscopic survey was to search for the
existence of a UCD population within the dynamically evolved Coma Cluster.
UCDs are extremely compact and faint objects having apparent magnitudes at the
distance of Coma in the
range $21.5 < V < 24.5$, colors up to 0.2 magnitudes redder than the red sequence,
and half-light radii corresponding to $0.01 - 0.2$ arcsec.
In ground-based
surveys, these objects are not resolved, but with the
resolution of ACS the larger ones are just resolved.  It is therefore
challenging, even with HST imaging, to distinguish these near point sources
from foreground stars.

To choose a sample of UCD candidates, we first identified point
sources in the catalog of \citet{adami} based on their large ground-based 
CFHT/CFH12K survey.  We then imposed the following criteria:
\begin{enumerate}
\item $R < 24$ ($3^{\prime\prime}$ aperture mags)
\item $0.45 < (B-V) < 1.1$ ($3^{\prime\prime}$ apertures)
\item Higher priority to sources with $0.15 < R-I < 0.6$
\item Located in regions with processed ACS images (available before 2007 February)
\item Gave highest weight to sources showing any sign of having 
profiles broader than stellar in our ACS images. 
\end{enumerate}
This last point biases the sample to larger objects although
unresolved sources were included in the target list as well.
The ($B-V$) range includes the Fornax/Virgo UCDs at the blue end and
compact ellipticals (cEs) at the red end.  In part because of 
potentially large photometric errors at these faint magnitudes,
we refrained from making our color cut too restrictive.  The final list contains
165 UCD candidates.  The weights were implemented by the Autoslit3 mask making 
software for choosing targets.

%--------------------------------------------------------------
\subsection{Observations}\label{obsspec}
%--------------------------------------------------------------

The spectroscopic setup was intended to maximize light throughput at
the expense of resolution.  As we only need to establish cluster
membership for these galaxies, and the Coma Cluster has a velocity
dispersion of $\sim1000$ km/s, a resolution of 200 km/s is  
adequate.   The spectral range was chosen to include the Ca H and
K lines and Balmer break at the blue end ($\sim 4080$\AA \ at the redshift
of Coma) and the Ca triplet at the red end (around 8600\AA). Because
the most critical spectral features for redshift measurements of
low redshift, faint galaxies are the Ca H \& K lines and Balmer break, 
we require high sensitivity in the blue.

In order to attain these specifications, we use the Keck LRIS low resolution 
imaging spectrometer \citep{lris95} in multi-object spectroscopy mode.  
A dichroic was used to split light
at 5600\AA \ between red and blue detectors.  On the blue-side, we used
the 400 line mm$^{-1}$ grism blazed at 3400\AA \ providing a dispersion of
1.09\AA \ pix$^{-1}$ and wavelength coverage of 4384\AA.   With 1.2 arcsec
slitlets, we achieved a resolution of $\sim$7.8\AA \ FWHM.
On the red-side we chose the 400 line mm$^{-1}$ grating blazed at 8500\AA \
with wavelength coverage 3950\AA \ and 1.92\AA \ pix$^{-1}$ dispersion.
These provide peak efficiencies of 50\% at 3900\AA \ and over 40\% at 7000\AA, 
along with the intended full wavelength coverage.

\begin{figure}[t]
\begin{centering}
\includegraphics[angle=90,totalheight=4.5in]{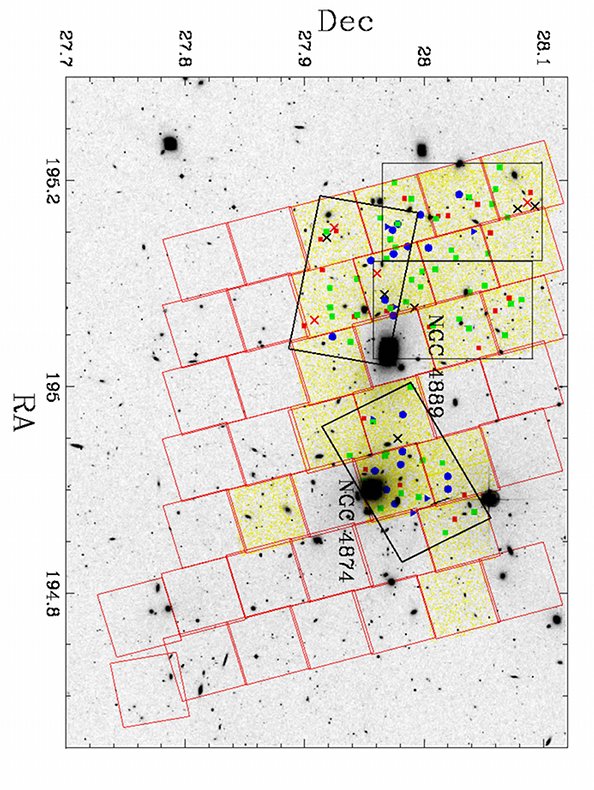}
%\plotone{fig4}
\caption[]{Central region of the Coma Cluster with overlay showing all
originally proposed ACS fields (highlighted yellow regions are completed ACS
imaging) and the 4 LRIS mask regions.  Green and red points represent
spectroscopically determined member and background galaxies
from our membership determination sample.
Blue circles and triangles are confirmed and questionable UCDs, while
red and black Xs represent UCD candidates which proved to be background galaxies
and stars respectively.
\label{map}}
\end{centering}
\end{figure}

We observed 4 masks over 2 nights 2-3 April 2008.  Details of these
observations are provided in Table \ref{obinfo}.
Each MOS mask covers a field of view of $5^{\prime} \times 8^{\prime}$.
Total integrations for each mask were $8 \times 1500$s with the exception
of the blue side of the 3rd mask for which we only obtained $6 \times 1500$s.
The masks covered 4 separate fields with
only slight overlap (4 objects were observed in 2 masks each),
and nearly cover the full core region with completed ACS 
imaging.  A total of 140 objects were targeted with the 4 LRIS masks
for an average of 35 slitlets per mask.  These included
20 expected background galaxies (membership probability 4),
73 low surface brightness probable/potential
cluster members (membership probabilities 1-3), and 47 UCD candidates.
In Figure \ref{map}, we overlay the locations of the
4 LRIS fields on an image of the central Coma Cluster region.  
Red boxes designate the originally proposed locations for the
ACS imaging; those highlighted in yellow have been completed.  The locations
of targets are also shown.

A subsequent observing run in 2009 focused on the continuation of our
UCD search.  UCD discoveries will be presented in a future paper.
However we did measure redshifts for two new LSB members.  We include
these measurements in Table \ref{taball}, but as these two
objects are not statistically relevant to the discussion in this paper,
we do not discuss them further.

Calibration arc lamps were observed each evening and sky lines were
used to determine observational shifts.  
Dome and halogen lamp flats were also observed daily.
The signal-to-noise per Angstrom for the galaxy spectra ranges from $\sim 60$ to less than
1. The brightest high surface brightness objects had S/N (around 5000\AA) $> 20$. 
Typical S/N for the low surface brightness dwarfs for which we
were able to measure redshifts ranged from $\sim 4-15$ around 5000\AA.  
For the UCD sample, membership was established for targets having
S/N $> 4.5$ while less secure redshifts were derived from spectra with
$2.5 <$ S/N $< 4.5$.

%--------------------------------------------------------------
\subsection{Data Reduction}\label{datared}
%--------------------------------------------------------------

Data were reduced using the standard procedures in IRAF.
We first overscan/bias corrected the images using tasks {\it lccdproc}/{\it lrisbias}.
For the blue-side data, 2 CCD chips read out with 4 amplifiers were corrected 
for different gains and 
tiled into a single image.  Data on the red-side were likewise corrected for
2-amp readout.
Halogen flats for each mask were combined and a normalized flat image was
generated with {\it apflatten}.  Science MOS images were divided by this
flat, and the 8 individual exposures for each mask were subsequently
median combined using sigma clipping after first fixing bad columns.

Slit spectra were rectified by first tracing the slit gaps for each mask and 
fitting these
with 4th order legendre polynomials.  {\it geomap} was run
to compute the 2D surface for the full set of slit gaps in a mask and {\it geotran}
was then run to
generate images with straightened slits using that transformation.
Once this operation was completed, individual 2-D spectra
were cut out.  Arc spectra were rectified and extracted for each 
slitlet in the same manner.  

We ran the IRAF tasks {\it identify} ({\it reidentify}) and {\it fitcoord} to wavelength
calibrate the lamp spectra, and {\it transform} to calibrate the object spectra
using these arc wavelength solutions.   Sky lines in the slitlet
spectrum of each object 
were used to correct for offsets from the lamp wavelength calibrations.
On the blue-side spectra,
the shift was calculated primarily with the $5577$\AA \ emission line.  However,
in many spectra this line was at the very edge of the spectrum or,
for objects with slits at the edge of the mask,
missing from the spectrum altogether.  Because we were unable to accurately
correct for the 
shift in many spectra, we believe
we may suffer from systematic errors as large as 100 km/s.
To extract 1-dimensional spectra, we used {\it apall} to identify the
spectrum center, width, and sky regions. The RVSAO package was used,
along with absorption and emission line template spectra, to measure 
cross correlated redshifts.

In the red-side spectra we had to contend with bright skylines
redward of 7000\AA.  Sky regions within the slitlet were used
to subtract these out, but residual noise in most cases is
brighter than the target galaxy signal.  We therefore only
used these spectra when the source signal was bright and to
search for emission lines.                                                                                

%-----------------------------------------------
\section{Results}\label{results}
%-----------------------------------------------

\begin{figure}[t]
\begin{centering}
%\plotone{fig5.eps}
\includegraphics[angle=0,totalheight=6.5in]{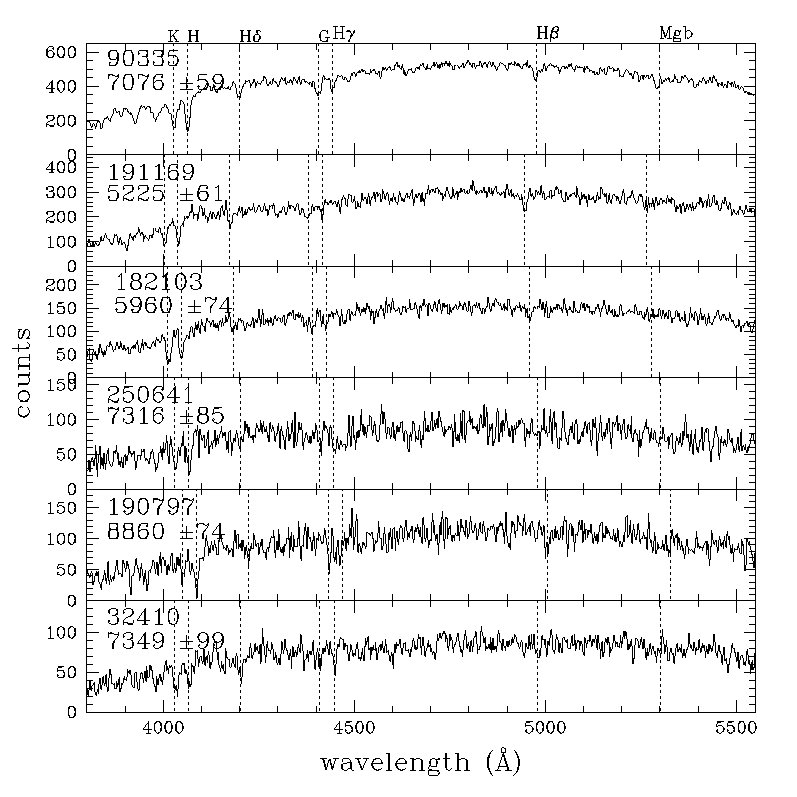}
%\plotone{fig5}
\caption[]{Spectra for 6 dE and dE,N, smoothed 3 times.  
From top to bottom, $R-$band magnitudes range from 19.3 to 21.8.
\label{specfig}}
\end{centering}
\end{figure}

In Fig. \ref{specfig}, we display 6 spectra for dE and dE,N galaxies from which 
we obtained redshift measurements, arranged in order of increasing
$R-$band magnitude.  
The majority of redshifts for Coma member galaxies were measured using prominent lines in the blue,
in particular the Ca H\&K, Balmer break, and H$\beta$ lines.  High redshifts 
were more often measured 
from emission lines in the red-side spectra.  Spectra from our
faint and low surface brightness dwarfs were most often too faint to
extract redshift measurements from our red-side spectra due to
bright sky emission. 

\begin{figure}[t]
\begin{centering}
%\plottwo{fig6a.eps}{fig6b.eps}
\plottwo{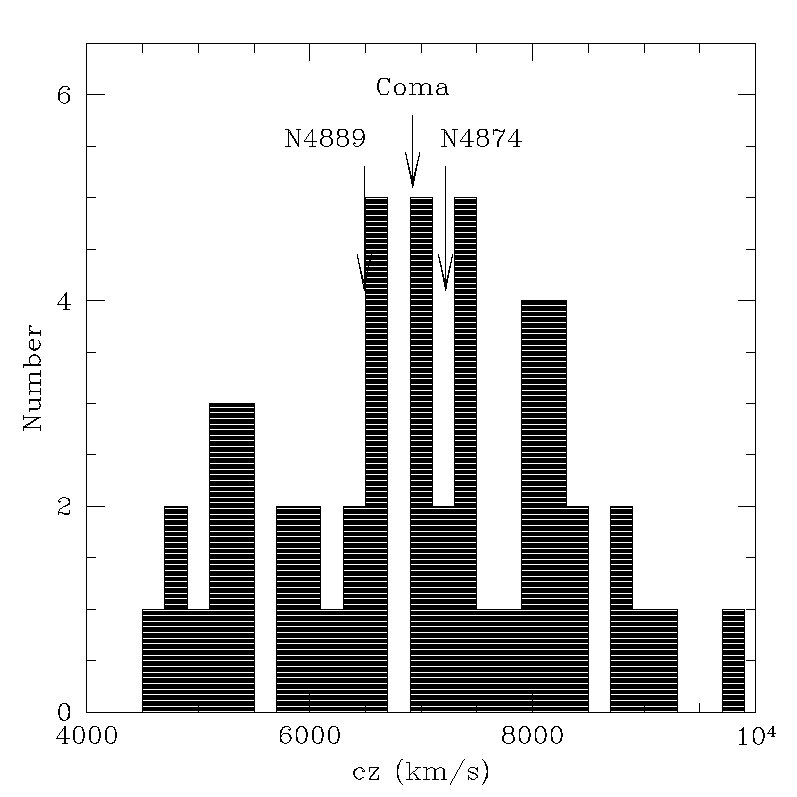}{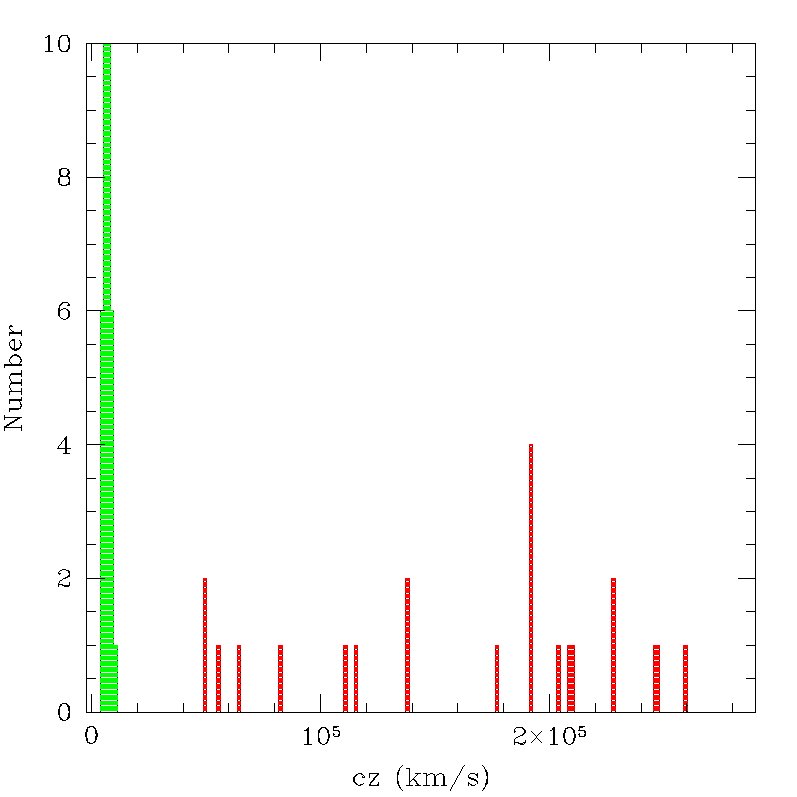}
\caption[]{Measured member and background redshifts. These do not include the 
UCD sample redshifts.
\label{velhists}}
\end{centering}
\end{figure}

The mean redshift of the Coma Cluster is cz $\sim 6925$ km s$^{-1}$ with
$\sigma \sim 1000$ km s$^{-1}$ \citep{wc99,gm01,shao94,ecbcmp02}.
We take Coma members to be those galaxies within $3\sigma$ of the cluster
mean velocity, or between $4000 - 10000$ km s$^{-1}$.
In total we measured redshifts for 70 Coma Cluster members including 19 UCDs.  
Spectra for the 19 objects confirmed as UCDs are provided in Chiboucas et al. (in prep). 
Another 14 objects have less secure redshift measurements consistent with Coma Cluster
membership including 6 potential UCDs.  Redshifts confirmed that another 24 objects are
background galaxies while 6 UCD 
candidates proved to be foreground stars.  For the remainder of the targets,
the S/N of the spectra was too low to extract redshifts.  In total, we 
measured redshifts for over 80\% of our targets.  
We present the redshift measurements in Table \ref{taball} for the non-UCD
sample and display histograms of the radial velocities in
Fig \ref{velhists}. 
Redshift measurements for the UCDs are provided in 
Chiboucas et al. (in prep).

\begin{figure}[t]
\begin{centering}
%\plotone{fig7.eps}
%\plotone{fig7}
\includegraphics[angle=0,totalheight=6.5in]{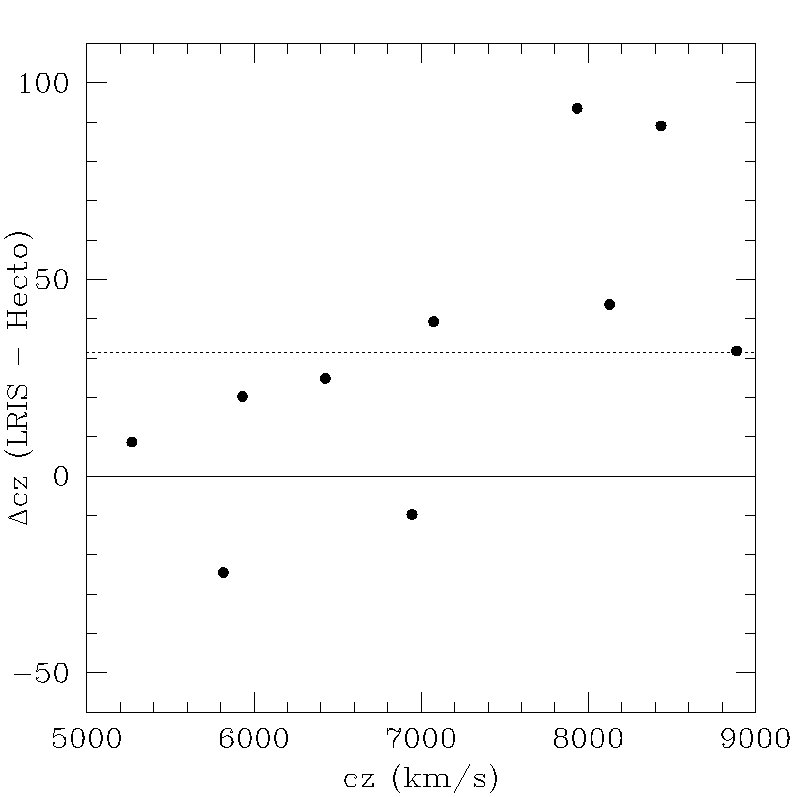}
\caption[]{Galaxy velocity measurements compared to those also measured
with Hectospec. The average difference is 32 km/s. 
\label{speccomp}}
\end{centering}
\end{figure}

In Figure \ref{speccomp} we compare radial velocity measurements for
galaxies that overlapped in the Hectospec sample. The average
difference for all 10 galaxies in common is 32 km s$^{-1}$.  Two 
objects differ by nearly 100 km s$^{-1}$.  These are objects which 
were observed on the far edge of LRIS masks cutting 
off much of the red portion of the spectra.  For these galaxies, and
up to 11\% of the sample, the blue-side spectra cut off at the red end by 
$\sim5300$\AA.  There are few
arc lines in this region, so the wavelength calibration is not well constrained at 
the red end,
and no sky lines from which to determine relative shifts from the arc 
calibration.  We therefore expect larger uncertainties for galaxies 
located along one side of the masks.   However, neither 
the small systematic offset from Hectospec velocities nor larger uncertainties
for some galaxies are significant compared to the Coma Cluster velocity
dispersion, and so do not affect the cluster membership determination.

For 51 non-UCD member galaxies, we find
$\langle v_r\rangle = 6970\pm178$ with $\sigma_v = 1269\pm126$.
Excluding the one highest redshift object at 9870 km/s, we find
$\langle v_r\rangle = 6912\pm172$ with $\sigma_v = 1213\pm122$.
\citet{ecbcmp02} found a mean radial velocity for the Coma Cluster
of 6925 km s$^{-1}$ with velocity dispersions
for the giant and dwarf galaxy populations of $979\pm30$ and $1096\pm45$
km s$^{-1}$ respectively.  The velocity dispersion found for  
our faint, low surface brightness sample is 1$\sigma$ larger than their
dwarf galaxy measurement.  This may be due in part to the smaller sample size
or smaller field coverage of this study, and in particular with the fact that
our sample is confined to the cluster core since velocity dispersion typically 
decreases with projected radius in clusters.
However, it is also consistent with the trend of increasing velocity dispersion 
with decreasing galaxy luminosity (or mass) previously found for the Coma 
cluster and which may be due to dynamical friction processes that lead 
to equipartition \citep{abm98}.

\begin{figure}[t]
\begin{centering}
%\plotone{fig13.eps}
%\plotone{fig13}
\includegraphics[angle=0,totalheight=6.5in]{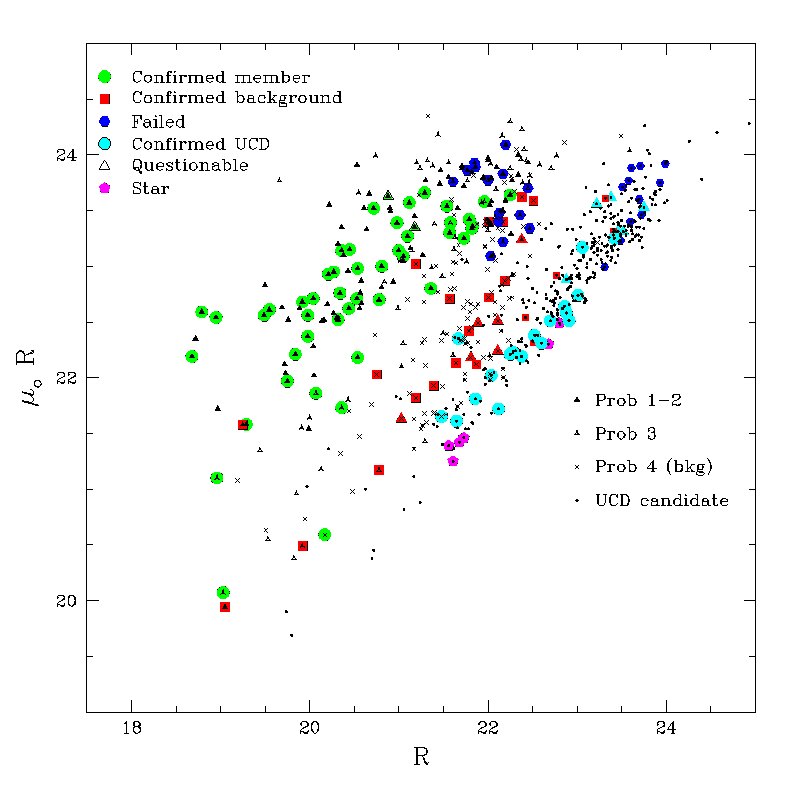}
\caption[]{Central surface brightness (mag arcsec$^{-2}$) vs. $R-$band magnitude.
Small symbols represent candidate targets and 
associated subjective membership probabilities as described in the key in the lower
right. Symbols enclosing these denote the object type and membership determination for
spectroscopically targeted objects (upper left key).  
Photometry comes from the catalog of \citet{adami}. Triangles refer to questionable 
cases where redshift measurements are uncertain due to lower S/N spectra.
\label{sb}}
\end{centering}
\end{figure}

\begin{figure}[t]
\begin{centering}
%\plotone{fig14.eps}
%\plotone{fig14}
\includegraphics[angle=0,totalheight=6.5in]{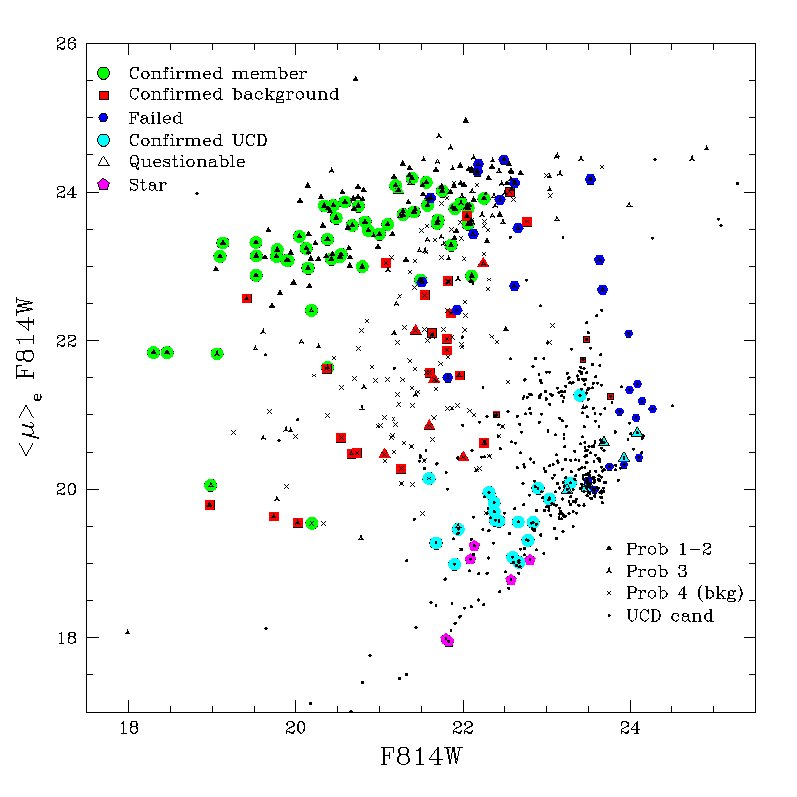}
\caption[]{Effective surface brightness (mag arcsec$^{-2}$) vs. $F814W-$band magnitude. 
Symbol types as in Fig. \ref{sb}.
\label{sbI}}
\end{centering}
\end{figure}

From the low S/N failures, we can determine our spectroscopic survey limits.
We find that in our 3.3 hr integrated exposures, we are able to measure 
redshifts for all targets with central surface brightness higher than $\mu_{\circ} (R) < 23$
mag arcsec$^{-2}$.  
For well resolved galaxies, we are able to measure redshifts down 
to $\mu_{\circ} (R) = 23.6$ 
for galaxies brighter than $R < 22$ (or to mean effective surface brightness 
$\langle\mu\rangle_{e,F814W} < 23.8$ for 
galaxies brighter than $F814W < 22$).  For the quasi-point source UCD candidates, we
reach a magnitude deeper, becoming incomplete at $R \sim 22.9$ and 
$F814W \sim 23.3$ (see Fig. \ref{sb}-\ref{sbI}).

\begin{figure}[t]
\begin{centering}
%\plotone{fig8.eps}
%\plotone{fig8}
\includegraphics[angle=0,totalheight=6.5in]{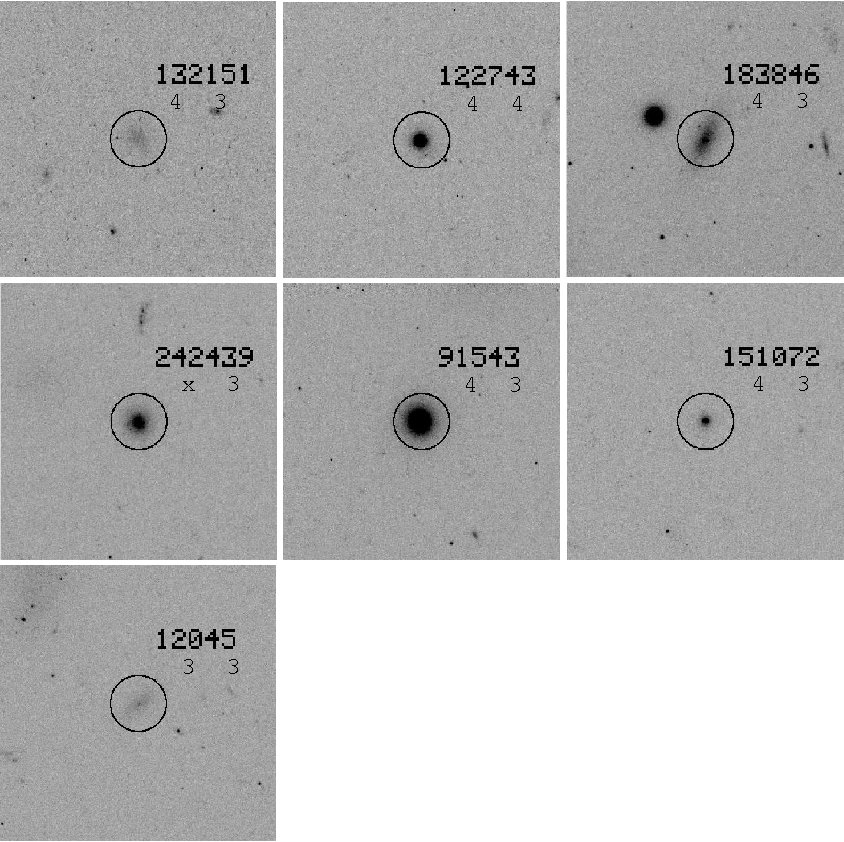}
\caption[]{Galaxies spectroscopically determined to be members which had
been assigned membership probabilities 3-4 (possible members or likely
background). The two membership assignments for each object are labeled.  For
object ID 242439, only one author supplied a membership assignment.
Images are 20 arcsec on a side.
\label{misid}}
\end{centering}
\end{figure}

From the redshift measurements of the 
sample selected to test membership probabilities we find a high
success rate for establishing membership through indirect means.
Of the 50 objects in our LSB sample that were confirmed to be members, 43 were 
expected to be members (probability classes 1-2) by one or both of
NT and HF.  Only 7 had been designated class 3-4 by both. 
We display thumbnails of these 7 morphologically ambiguous objects in Figure \ref{misid}.
All 7 are classified as types other than dE.  One is classified as an irregular, 4 as 
normal ellipticals, and 2 as S0 and dS0.  This last was given the membership
probability 3 (uncertain) by both authors.  Based on the spectroscopic 
redshifts, we see that two of the confirmed members classified as E 
types (91543, 122743) are in fact compact elliptical (cE) galaxies.  The other two galaxies 
classified as Es are even more compact
high surface brightness objects. One (151072) may be a member of the UCD
class based on its size and surface brightness.  The other (242439) is larger and 
appears to have a low surface brightness envelope with some additional structure.
The objects correctly identified as members based on 
morphology and surface brightness are largely of the dE and dE,N classes with only a few 
exceptions. The failures are largely featureless, and often round, compact
early types.  

\begin{figure}
\begin{centering}
%\plotone{fig9.eps}
%\plotone{fig9}
\includegraphics[angle=0,totalheight=2.0in]{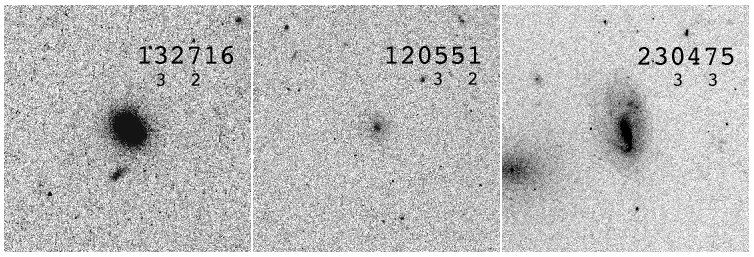}
\caption[]{Three galaxies with redshifts placing them outside of the Coma Cluster.
Membership classifications based on morphology from 2 authors are listed under 
the galaxy ID.  None
of these 3 galaxies were considered a definite background galaxy.  Images are
20 arcsec on a side.
\label{misid2}}
\end{centering}
\end{figure}

Of the 20 confirmed
background objects, 17 had been correctly assigned a 4 (definite background) by
at least one author.  The remaining 3 had been classified as 3 (uncertain)
by at least one author.  We show these 3 objects in Figure \ref{misid2}.   It
is easy to understand the confusion of two of these objects with cE and
dE,N type cluster galaxies.

\begin{figure}[t]
\begin{centering}
%\plotone{fig10.eps}
%\plotone{fig10}
\includegraphics[angle=0,totalheight=6.5in]{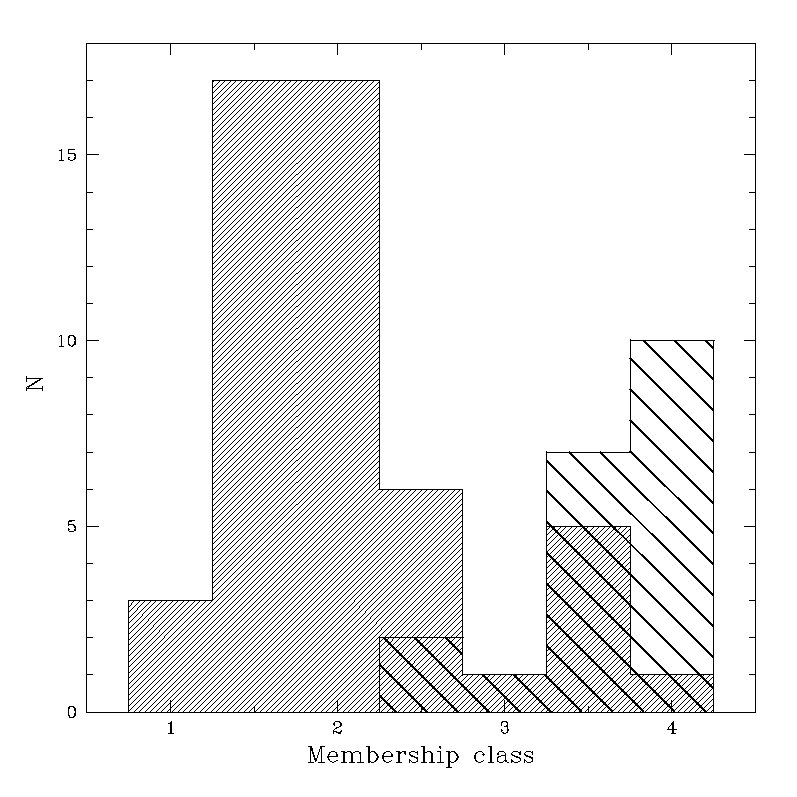}
\caption[]{Histogram of membership probability results.  The shaded
and open hatched histograms display confirmed member and background galaxy counts
respectively as a function of average membership probability designation.
\label{classhist}}
\end{centering}
\end{figure}

These results indicate very good success
for establishing membership through indirect means, although failures are clearly
due to morphological biases inherent in using this methodology. 
We graphically display these results in Fig. \ref{classhist}.
The shaded histogram shows, as a function of the average membership probability,
the number of objects confirmed to be members while the open hatched histogram shows
counts for confirmed background galaxies.  For objects designated
as 1-2 by both authors, we find a 100\% success rate.  For objects 
classified as 4 (background) by both, we find a 91\% success rate.  We note that
our small background sample does not reflect the full range of likely background galaxies. 
The great majority of background galaxies have spiral arms and other tell-tale features and
can be confidently associated with the background.  We expect this 91\% success rate to therefore
be a lower limit.  Between the expected member and background galaxies,
in the range where at least one author designated the galaxy as
an uncertain 3, we encounter the problem cases.  Objects with
an average probability of 2.5 turned out to be members 75\% of
the time, those with a probability of 3 turned out to be members
half of the time, and those with a probability of 3.5 turned out
to be members 42\% of the time.  On average, for these 22 cases where
at least one author chose the uncertain probability 3, 55\% turned
out to be members while 45\% turned out to be background.  This
is very close to a 50-50 split one might expect for such an uncertain
category.  For the full sample of morphologically classified galaxies in our ACS fields, this would
imply that $\sim90$\% of 400 galaxies ranked $1-2.5$ may be expected to be members
and $> 70$\% of 364 galaxies designated $3.5-4$ may be expected to lie in the background.
Another $\sim150$ galaxies from the uncertain category 3.0 may also prove to be members.

In summary, we find that if mistakes are made with membership assignments based on morphology, it
is often because a small fraction of textureless high surface brightness objects 
taken to be in the background are compact dwarfs in the cluster.   Candidates expected to
be members almost always are.  Candidates in the uncertain category split almost
evenly between member and non-member.
 
For the UCD sample, our spectroscopy measurements become incomplete
by $R = 23.3$.  For targets brighter than this, where we are able to measure
secure redshifts, an astonishing 19, or 66\%, have turned out to be members.  Only
4 targets were found to lie in the background while 6 proved to be stars.  We
discuss these results in greater detail in Chiboucas et al. (in prep).  

\begin{figure}[t]
\begin{centering}
%\plotone{fig11.eps}
%\plotone{fig11}
\includegraphics[angle=0,totalheight=6.5in]{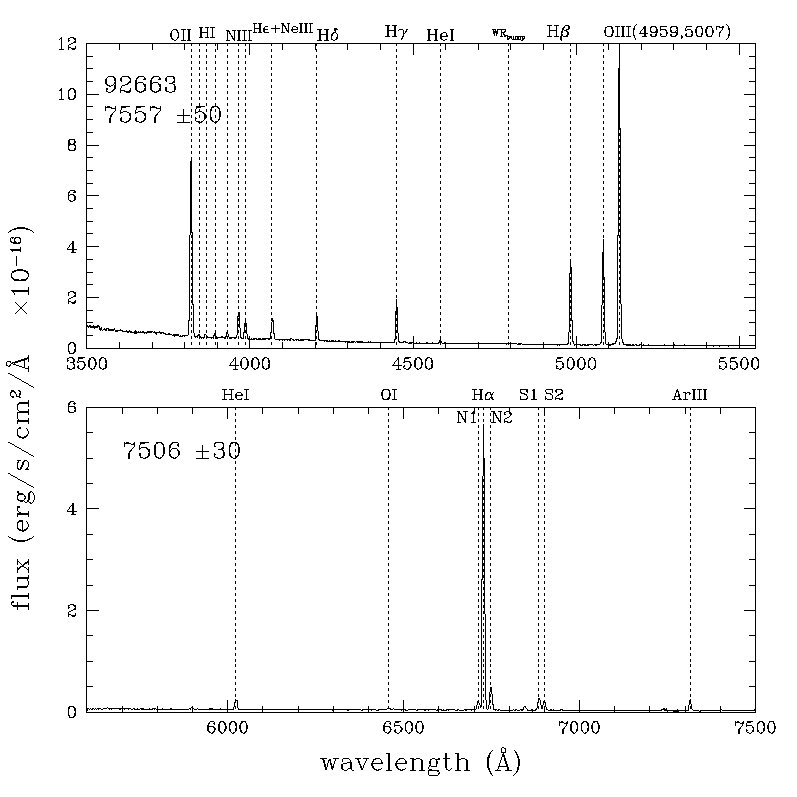}
\caption[]{Relative flux calibrated blue and red-side spectra of an 
unusually strong emission line object.
\label{espect}}
\end{centering}
\end{figure}

One object (92663) targeted originally as a UCD candidate was confirmed to be a cluster member
with $v_r = 7525$ km s$^{-1}$ through unusually strong emission lines. 
We show the relative flux calibrated red and blue-side spectra in Fig. \ref{espect}.
The spectrum exhibits strong emission lines along with
continuum. No absorption lines are apparent in this spectrum.
Nebular He and a Wolf-Rayet bump centered around restframe $4660\AA$ are also visible,
indicative of a young starburst.
Placing our line strength measurements of $\log \ $[O III]5007/H$\beta$ (0.6) and $\log \ $[N II]6583/H$\alpha$
(-1.0) on a BPT diagram \citep{bpt81}, we find our object lies on the branch 
of objects that are strongly star forming/HII regions.  Because these are line ratios of 
neighboring lines, effects of reddening and flux calibration uncertainties are minimized.  
From [NII]$\lambda6583 /$ H$\alpha$ (the N2 index) we calculate the oxygen 
abundance 12 + log(O/H) = 8.3 using the relation found by \citet{pp04}, somewhat
lower than the solar value 8.66.  
We measure an equivalent width of $1008\AA$ for H$\alpha$ well above the threshold
for the definition of a starburst \citep{lee09}.  Using STARBURST99 models \citep{leitherer99},
we estimate an age of this starburst of $3.1 - 4.1$ Myr depending on the model parameters
used, consistent with the Wolf-Rayet features observed in the spectrum.
As might be expected from the strong H$\alpha$ emission line, this object has been detected in 
both NUV and FUV with GALEX along with H$\alpha$ imaging from the 2.5m Isaac Newton Telescope.
These data are discussed further in \citet{smithHa}.

\begin{figure}[t]
\begin{centering}
%\plottwo{fig12a}{fig12b}
\includegraphics[angle=0, scale=0.45]{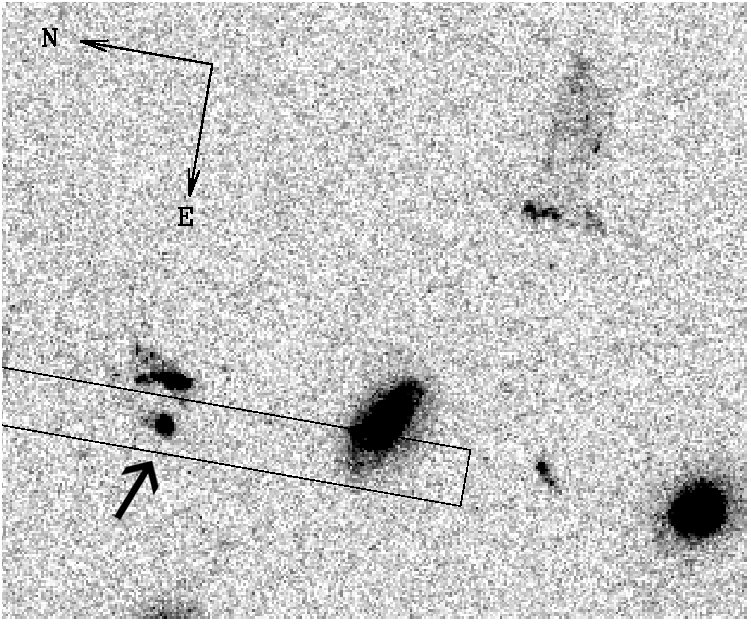}
\includegraphics[angle=0, scale=0.45]{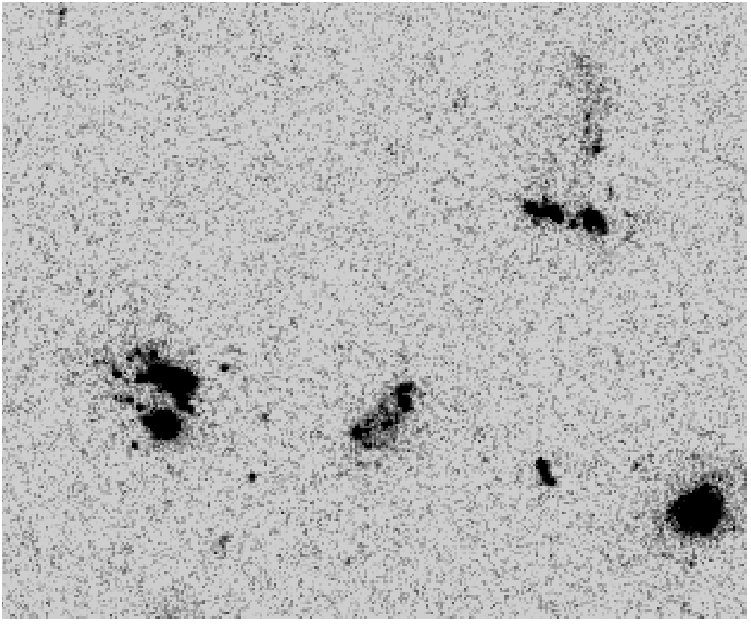}
\caption[]{Candidate UCD emission line object. Images are 15.5 arcsec 
across.  Top: The F814W
image with the target object (indicated by an arrow) centered in the slit. Another object lies just
outside the slit to the west.  The other galaxy
further along this slit is also observed.    Bottom: The same section of
the F475W image.    
\label{slit}}
\end{centering}
\end{figure}

This object, targeted as a UCD candidate, has a magnitude and size in the $F814W$ 
band expected for UCDs.  Just to the west of this object, and just outside
the slit, is another object visible in the $F814W$ image with a high surface
brightness core and distorted envelope (Fig. \ref{slit}).
However, from the $F475W$ band image, we find these two objects are very blue
and more extended than seen in the $F814W$ image. They are
most likely associated, perhaps as an interacting pair or a single dwarf
irregular galaxy having two separate star bursting regions.  The targeted object
is most likely a large HII/OB complex.
The magnitude of the targeted object
is $F814W = 23.1$, the one just above is a half magnitude brighter. Both objects
are quite blue: the targeted source has $F475W - F814W = -0.93$ while
the second object has $F475W - F814W = -0.37$. 
The double object spans $\sim2.5$ arcsec, or at the distance of Coma, 1.1 kpc.

We note that there is another unusual blue object in the immediate vicinity which appears
to be a tidal stream or tidally disturbed galaxy with brighter patches at one end having,
presumably, high star formation rates. If this object is also a Coma Cluster member,
these two star bursting objects would be separated by a projected distance
of $\sim4.4$ kpc.  It would seem unusual to find such strongly star forming objects
in an old, evolved cluster.  
However, similar objects have been found in the Coma core
region.  \citet{fb08} find an unusual starbursting complex 
to the southwest of the core which they refer to as 'fireballs' due to the multiple
blue blobs and streams associated with this object.  The fireballs appear to be
associated with a nearby merger remnant galaxy.
Near the 'fireballs', \citet{yagi07} find a long H$\alpha$ emitting stream 
associated with a post-starburst galaxy.  
In a study of Coma supercluster galaxies, \citet{maha10} find that star-forming dwarf
galaxies on first infall through the dense intracluster medium undergo a burst of star formation 
and are subsequently quenched.  This may be what we are seeing here, although tidal 
effects could also play a role.
This emission line object is discussed further in \citet{smithHa}.

%-----------------------------------------------------------
\section{Discussion}\label{disc}
%--------------------------------------------------------

We display membership results in magnitude - surface brightness space
in Figures \ref{sb} - \ref{sbI}.  $R$-band photometry comes from the ground-based
\citet{adami} catalog. $F814W$ photometry comes from SExtractor measurements from
our ACS data \citep{hammer10}.     
Underlying symbols denote the original membership
probability.  Those highlighted in green circles were confirmed through redshifts
to be members, while red squares highlight the background galaxies.  Cyan
circles represent confirmed UCDs.   Two roughly parallel sequences
of cluster members are visible in these figures.  The normal dwarf population lies along
the low surface brightness end of the range and follows the well known magnitude -
surface brightness relation for dwarf ellipticals.  On the other
side of the surface brightness spectrum are the very high surface brightness and compact
UCDs which appear to follow a sequence parallel to that of the dEs.  
This region is closely bounded by the stellar sequence.
Note that due to seeing limitations, the ground based photometry generates
narrow UCD and stellar sequences that merge and are bounded on the high
surface brightness side by the seeing disk. Although there is still some overlap,
ACS measurements 
expand this surface brightness space and do a better job at separating
UCDs and stars.  In between the compact and LSB sequences,
the great majority of objects are confirmed and expected background
galaxies.  

These figures demonstrate that, at least in the case of normal dwarf
galaxies, surface brightness is a good, and reliable, indicator of
membership.  In the lowest surface brightness regime,
all objects with measured redshifts are members.  This result is significant
because our spectroscopic survey becomes incomplete due to low surface
brightness before becoming incomplete at faint magnitudes.  From these diagrams,
we may assume that lower surface brightness galaxies without spectra, brightward
of $F814W \sim 21.5$, are members as expected from their membership probability
classes.  

However, these figures also illuminate several issues with this indirect membership
assignment technique.   At intermediate surface brightnesses, member galaxies
overlap in magnitude and surface brightness with the background population.
In this region we will expect
greater uncertainty with the membership assignments.  Also evident
is the surface brightness bimodality in the Coma Cluster member 
population.  At the highest surface brightnesses, there is a second
significant population of UCD cluster members  
which overlap, even in ACS images, with the stellar sequence.
The UCD sequence may continue
to brighter magnitudes than we find here, but in our 2008 observing run we did 
not target any UCD candidates brighter than $R = 21.5$. 
In addition there are a pair of cEs that fall within the surface brightness-magnitude
region of the background population.
It is strikingly apparent that if these two classes of compact objects constitute
significant populations in the Coma Cluster, we may be seriously mis-representing
the Coma Cluster galaxy population and associated properties in our
comprehensive cluster study when using surface brightness criteria to 
assign membership.  

A number of questions must be addressed.
We need to determine if compact dwarf ellipticals compose a significant fraction 
of the cluster population,
and how they might be identified using other means than spectroscopy.  
This is addressed in \citet{compact}.  With a total of 7 confirmed
members at present, it does not appear that compact dwarf galaxies form a large 
Coma Cluster population.  Similarly, we need to determine whether UCDs 
contribute significantly to the overall cluster dwarf
population, how they might be identified short of all-object spectroscopic surveys, 
and, most importantly, what
type of object these are.   We address these questions in detail in Chiboucas et al. (in prep). 
A $\sim$60\% success rate with our initial observations indicates a large
population. 
However, we do not yet know whether these
objects are the stripped remains of nucleated dwarf ellipticals or
simply giant star clusters, perhaps forming the bright tail of the 
globular cluster luminosity distribution.   

\begin{figure}[t]
\begin{centering}
%\plottwo{fig15a}{fig15b}
\includegraphics[angle=0,totalheight=3.5in]{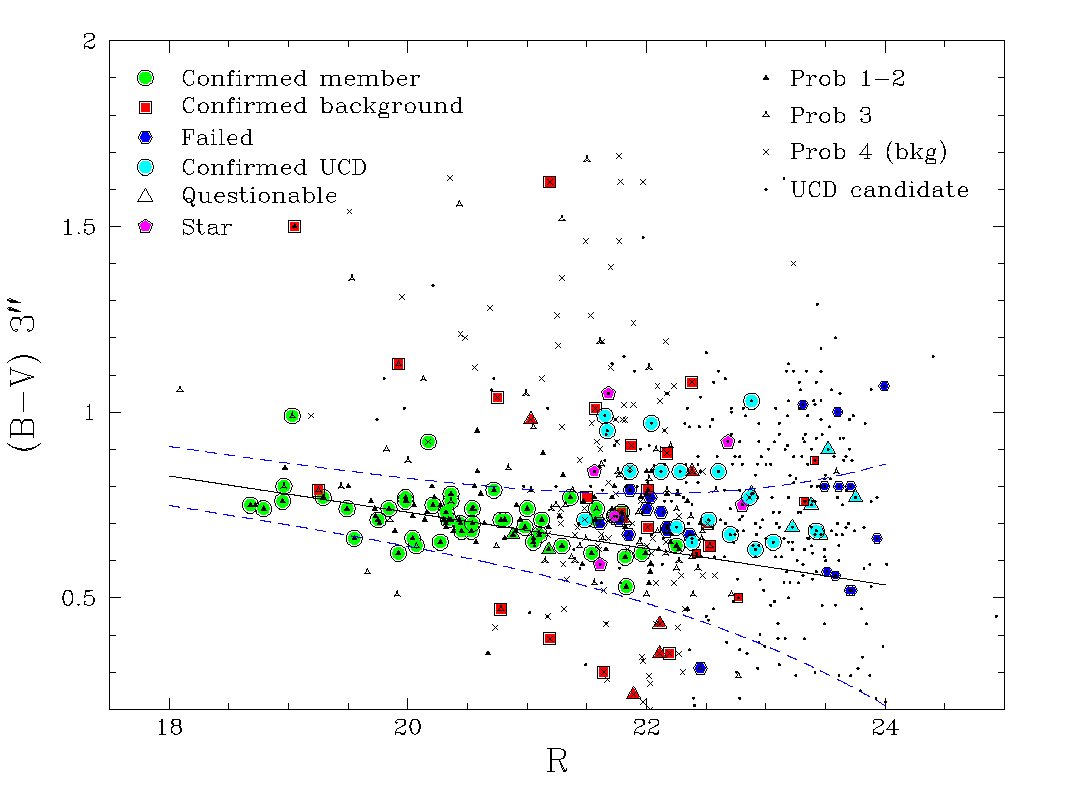}
\includegraphics[angle=0,totalheight=3.5in]{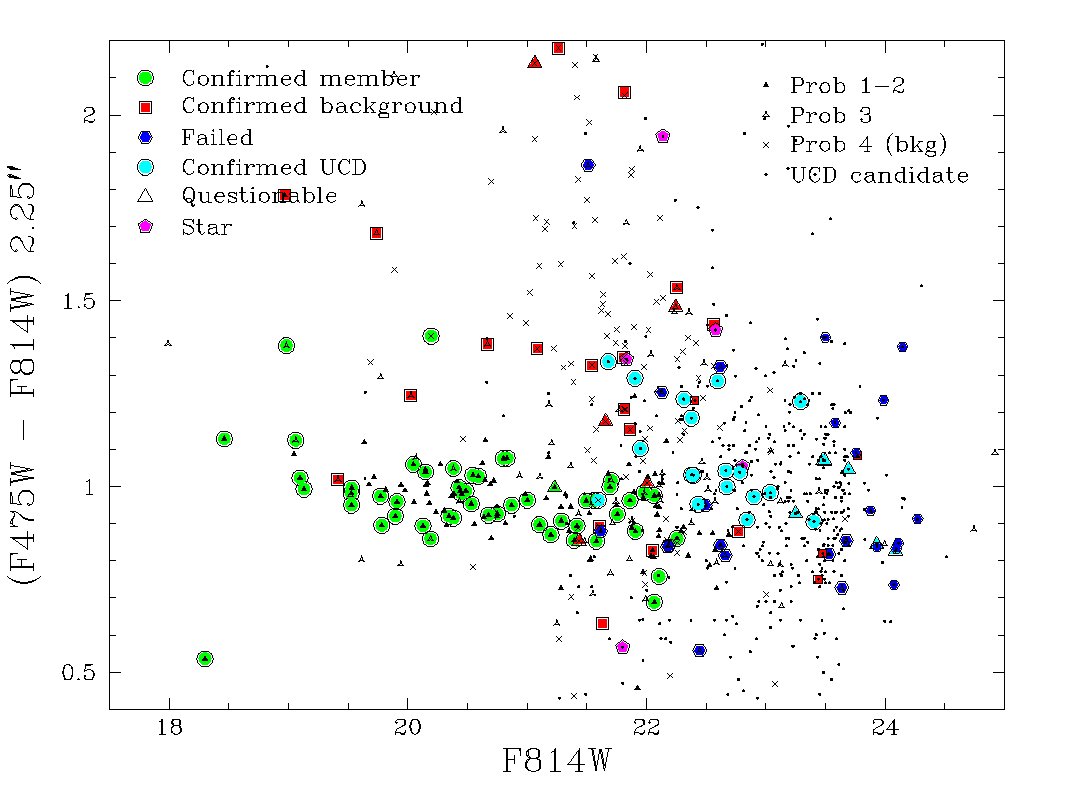}
\caption[]{Aperture color vs. total magnitude. Symbols as in the previous figures.
The solid line in the top plot represents the
best fit to 120 red sequence Coma Cluster members with $14 < R < 19$ 
extended to faint magnitudes.  The $\pm1\sigma$ error curves for the fit to brighter galaxies, 
including the contribution from photometric measurement errors, are also shown.
\label{colacs}}
\end{centering}
\end{figure}

We consider other means of identifying cluster members.  In our original classification,
color information was not used.  We display color-magnitude diagrams
in Figure \ref{colacs}.  The red sequence of early type cluster members
is immediately apparent (highlighted green circles).  
We include the best fit to the red sequence of 120 $14 < R < 19$ galaxies extended to
the faint magnitudes here and find that the red sequence is continuous down to at least 
$R < 22$. 
Brightward of $R$ or $F814W = 21.5 (M_{R/F814W} = -13.5)$, we expect that color 
can be used to distinguish members 
from background galaxies with high success.  There are only three discrepant 
points.  One background galaxy, originally classified as an uncertain membership
probability 3 falls right along the red sequence.  Incorporating colors
in the membership determination would most likely have led to the
mis-identification of this galaxy as a cluster member.  Two members
lie 0.2 magnitudes redward of the red sequence. These are
the already mis-identified cE galaxies.  Including color information for these
may reinforce the mis-identification as background galaxies.  
This study suggests that compact ellipticals with $B-V$ colors $\sim 0.2$ magnitudes 
redward of the red sequence warrant spectroscopic attention.
We also find that the confirmed UCDs have a large
and unexpected color spread.  There appears to be a subset that 
may follow the faint extension of the red sequence while the remainder of UCDs
have extraordinarily
red colors.   Thus, although we would expect that incorporating color 
information will 
improve our overall membership assignment success rate, especially 
for galaxies in the uncertain probability 3 category, and can certainly be used
to reliably exclude very red high redshift galaxies, it fails for some of 
the same objects as morphology and surface brightness criteria.  Other
means to differentiate between background and cluster members, such as the 
use of structural parameters, are being investigated.

The observed bimodal distribution of member surface brightnesses 
therefore complicates membership determinations.
We find that our membership assignments are largely successful to 
$R = 21.5$.  We have not yet uncovered a UCD population brighter
than this.  Our largest source of uncertainty in this range would
appear to be due to the cE class.  If these objects do not comprise
a large galaxy population, we may expect to have nearly complete
and accurate membership information down to $M_R = -13.5$.  We are,
however, continuing to investigate this in future work,
including new LRIS and Hectospec observations specifically designed to search
for bright UCDs and faint cEs.   

\citet{ma08} establish membership for Coma Cluster galaxies using a very
similar method.  Based on morphology, surface brightness, and size criteria 
they identify members in CFHT/CFH12K $B$ and $V$ imaging.   They found good
agreement with literature redshift results down to a completeness limit
of $M_B = -15$ ($B = 20$), or, assuming $B-R \sim 1.2$, to $R \sim 18.8 \ (M_R = -16.2)$.
At the time, spectroscopic redshifts were nonexistent for most galaxies fainter
than this.  
We compare our results and find 19 objects in common.  Of 2 spectroscopically 
confirmed non-member galaxies, they correctly identified both as background.
Of 17 confirmed members, they correctly assign 9 as probable members,
while incorrectly classifying 8 as distant spirals.  Our membership
probabilities for these same objects based on ACS images correctly assigned 
16 of the 17 members 
as definite to likely members although also incorrectly assign one of the two 
background objects 
as a likely member.  Overall, our membership assignments do a better
job at these faint magnitudes and surface brightnesses due to the much
higher spatial resolution of the HST/ACS images.  Since the success of morphological
memberships is directly related to the spatial resolution of the imaging
this method can presently only be used for very nearby groups and clusters; at the distance of
the Coma Cluster, spaced-based imaging becomes necessary.
At greater distances, it becomes difficult to use this method to probe 
to the same limiting magnitudes ($M_R = -12$) as this study.  Galaxies become fainter and 
smaller, more cluster galaxies with brighter intrinsic magnitudes will not have spectroscopic 
redshifts, and evolutionary effects may become a factor.  
We expect this method to be useful using HST/ACS images for clusters and groups within $\sim100$ Mpc. 

%-----------------------------------------------------------
\section{Summary and Conclusions}\label{conc}
%--------------------------------------------------------

In summary, we have used spectroscopic redshifts to test the success rate
for establishing Coma Cluster membership through indirect means such as
morphology, surface brightness, size, and color criteria.   
Using Keck/LRIS multi-object
spectroscopy, we targeted 93 galaxies to test membership probability assignments
using these indirect means, and another 47 compact objects to search for
UCDs.   

The results
of the membership tests showed that we can use surface brightness, size, and
morphology criteria, when based on high resolution ACS images, with good success.
We find a high success rate, with 100\% of galaxies that were
expected to be members proving to be members, and over 90\% of galaxies expected to
lie in the background proving to be background objects.   In practice, we expect
this latter percentage to be higher, since the
vast majority of expected background galaxies are those which 
exhibit structural features like spiral arms and these can be confidently 
associated with the background.  Our small sample of expected background galaxies
does not reflect the full background population. 
It is only between these expected member and expected background
categories, where galaxy memberships were assigned as 'uncertain' possible 
members, do we experience 
a breakdown with this method.   In these cases the failure rate for correctly
assigning objects as (possible) members is about 50\%, as
might be expected.  These are primarily cases of galaxies with  
surface brightness intermediate between low surface brightness dEs and higher
surface brightness background galaxies.  

In addition to these
normal dwarf galaxies, we confirm 19 Coma Cluster UCDs with magnitudes in the
range $21.5 < R < 23.5$ ($-13.5 < R < -11.5$) and having a large spread in
color which ranges from the expected color at the faint extension of the red 
sequence to 0.4 magnitudes redder than the red sequence.  We
discuss this sample further in Chiboucas et al. (in prep). 

The main conclusions from this study are as follows.  
The dominant cluster population, dwarf elliptical galaxies, follow a 
magnitude - surface brightness relation (which exhibits some scatter in surface
brightness) in
which dwarfs decrease in surface brightness at fainter magnitudes with little change in
size.  This property allows for the use of surface brightness and size as cluster membership
indicators for dwarf galaxies.
From our spectroscopic results we determine that very low surface brightness galaxies 
can indeed be confidently associated with the cluster.  It is only at the
higher surface brightness side of the magnitude - surface brightness relation
that some ambiguity enters into the membership 
determination as members and
background galaxies begin to merge and membership assignments become uncertain.
For statistical purposes, we find that objects with uncertain membership assignments
may be considered members with $\sim50$\% probability. It is only near the 
intermediate surface brightness boundary between members and non-members 
where this will be a problem; the majority of dwarf galaxies have
lower surface brightnesses and can safely be assigned as cluster members.

That said, we find that dwarf cluster members are clearly separated into LSB
and the compact cE and UCD families.  
The Coma Cluster galaxy population consists of the established low surface 
brightness dwarf galaxies and, on the other side
of the surface brightness spectrum falling far off the 
magnitude - surface brightness
relation of dE galaxies, are the very high surface brightness UCDs.
In between, largely within the range of the background population, are the cEs.
It is possible that UCDs and cEs form a single family of objects; 
the two-magnitude gap between these two populations at $-15 < M_{R} < -13$ must be explored.
Membership criteria which stipulate low surface brightness will fail
entirely for these compact galaxy types.

We did not use colors in our initial membership probability assignments, but
we investigate the membership results in terms of galaxy color.  For the most
part, we find colors do at least as well as surface brightness and size criteria
in distinguishing member from non-member galaxies.  Most cluster dwarfs follow
an obvious red sequence. 
However, color criteria fail for many of the same objects, namely the cE and 
UCD classes, where other indirect means fail.
Thus, colors provide a good discriminant most of the time but must be lenient on 
the red side to catch the cE and UCD components of the cluster population.  
Spectroscopy is required to establish membership for
these object types.

Overall, we find good success using morphology and structural characteristics based
on high resolution ACS imaging to establish membership,
with shortfalls at the boundaries between member and non-member population
properties, and in the high surface brightness regime.
Construction of a luminosity function may be problematic if cEs and
UCDs comprise significant cluster populations.
However, we do not believe cEs to be a dominant population and it is still unclear
as to whether UCDs are, in fact, galaxies at all rather than giant star clusters.
We are following up on these two populations in \citet{compact} and Chiboucas et al.
(in prep).  

Our results show that indirect and direct (spectroscopic) means for 
determining cluster membership have orthogonal biases.  Spectroscopy is
most successful for high surface brightness objects, the exact region where
the indirect means tend to fail.  Observations for high surface brightness
objects require much shorter exposure times so larger samples of high surface 
brightness galaxies may be observed in order to establish membership, while indirect 
means are expected to successfully identify LSB members.  This finding is significant as
it suggests ways to maximize the efficiency of telescope time for future 
cluster membership studies.  With high spatial resolution imaging, indirect
means provide a nearly 100\% success rate for identifying low surface brightness
members; spectroscopic studies should concentrate on intermediate and high surface 
brightness galaxies.

%--------------------------------------------------
\acknowledgments
%------------------------------------------------
We would like to thank the anonymous referee for helpful suggestions that have improved
this paper.
Based on observations made with the NASA/ESA Hubble Space Telescope, obtained at the Space Telescope 
Science Institute, which is operated by the Association of Universities for Research in Astronomy, Inc., 
under NASA contract NAS 5-26555. These observations are associated with program GO10861.
Support for program GO10861 was provided by NASA through a grant from the Space Telescope Science Institute, 
which is operated by the Association of Universities for Research in Astronomy, Inc., under NASA 
contract NAS 5-26555.  Some of the data presented herein were obtained at the W.M. Keck Observatory, which 
is operated as a scientific partnership among the California Institute of Technology, the University of 
California and the National Aeronautics and Space Administration. The Observatory was made possible 
by the generous financial support of the W.M. Keck Foundation.
The authors wish to recognize and acknowledge the very significant cultural role and reverence 
that the summit of Mauna Kea has always had within the indigenous Hawaiian community.  We are 
most fortunate to have the opportunity to conduct observations from this mountain.

\bibliographystyle{apj}
\bibliography{kc}

%------------------------------------------------
% TABLES
%------------------------------------------------
\clearpage
\begin{deluxetable}{ccllrrr}
\tabletypesize{\scriptsize}
\tablewidth{0pt}
\tablecaption{Keck/LRIS Observations \label{obinfo}}
%\tablenum{1}
\tablehead{
\colhead{Mask} &
\colhead{Date} &
\colhead{$\alpha$} &
\colhead{$\delta$ (J2000.0)} &
\colhead{PA (deg)} &
\colhead{Seeing (arcsec)} &
\colhead{N$_{slit}$} 
}

\startdata

1 & 2 Apr 2008 & 13 00 40.69 & 28 01 54.86 & 1.5 & 0.7 &  35 \\
2 & 3 Apr 2008 & 13 00 17.82 & 28 01 26.81 & 1.5 & 1.0 &  32 \\
3\tablenotemark{\dagger} & 3 Apr 2008 & 13 00 24.62 & 27 56 26.11 & 80.0 & 1.0 &  38 \\
4 & 2 Apr 2008 & 12 59 40.10 & 27 59 06.28 & 121.0 & 0.8 &  39 \\

\enddata

\tablenotetext{\dagger}{Exposure times for both red and blue chips were $8\times1500$ sec with the 
exception of Mask 3 for which we obtained only $6\times1500$ sec on the blue side.}

\end{deluxetable}

%\end{document}

\begin{deluxetable}{rccrrrrrrrrrrrrrrc}
\rotate{}
\tabletypesize{\scriptsize}
\tablewidth{0pt}
\tablecaption{LRIS (Membership test sample) redshift results \label{taball}}
%\tablenum{1}
\tablehead{
\colhead{ID} &
\colhead{T/F\tablenotemark{a}} &
\colhead{$R$\tablenotemark{b}} &
\colhead{$F814W$} &
\colhead{$F814W$} &
\colhead{$B-V$\tablenotemark{b}} &
\colhead{$g-I$} &
\colhead{$\mu_{\circ}$\tablenotemark{b}} &
\colhead{$\langle\mu\rangle_{e}$} &
\colhead{RA} &
\colhead{Dec} &
%\colhead{ACS} &
%\colhead{x} &
%\colhead{y} &
\colhead{cz} &
\colhead{err\tablenotemark{c}} &
\colhead{R$_{fx}$} &
\colhead{cz} &
\colhead{err} &
\colhead{SNR\tablenotemark{d}} &
%\colhead{R$_{fx}$} &
\colhead{type} \\
\colhead{} &
\colhead{prob} &
\colhead{} &
\colhead{corr.} &
\colhead{} &
\colhead{$(3^{\prime\prime})$} &
\colhead{$(2.25^{\prime\prime})$} &
\colhead{$(R)$} &
\colhead{$(F814W)$} &
\colhead{(J2000.0)} &
\colhead{} &
%\colhead{fld} &
%\colhead{} &
%\colhead{} &
\colhead{Ab} &
\colhead{km s$^{-1}$} &
\colhead{} &
\colhead{Em} &
\colhead{km s$^{-1}$} &
\colhead{} &
%\colhead{} &
\colhead{} 
}

\startdata

\bf{Members} &   & & & & & & & & & &  & & & & &  \\
132151 &  4  3 & 21.58 & 21.43 & 21.70 & 0.74 & 1.02 & 23.39 & 23.62 & 12 59 26.04 & 28 01 04.00 &  5486 & 130 & 2.9 & & & 6.3 &   ImV \\
133494 &  2  1 &       & 18.24 & 18.46 &      & 1.13 &       & 21.84 & 12 59 30.83 & 28 02 30.73 &  6549 &  77 & 8.5 & & & 44.6&    E \\
190797 &  3  2 & 21.79 & 21.20 & 21.58 & 0.73 & 0.85 & 23.42 & 23.82 & 12 59 31.73 & 27 57 48.17 &  8860 & 74 & 4.0 & & &  5.0   &  dE \\
191169 &  2  1 & 19.98 & 19.82 & 20.13 & 0.76 & 0.89 & 22.56 & 23.24 & 12 59 34.35 & 27 59 43.41 &  5225 & 61 &  9.0 & & & 15.3  &  dE \\
192254 &  3  1 & 22.25 & 21.80 & 22.25 & 0.64 & 0.86 & 23.64 & 23.91 & 12 59 35.23 & 27 58 50.28 &  7355 & 110 &  3.1 & & & 4.5 &  dE \\
194749 &  3  1 & 21.73 & 21.54 & 21.86 & 0.72 & 0.96 & 23.25 & 23.28 & 12 59 38.08 & 27 56 56.45 &  6128 & 79 &  3.8 & & & 7.1 &  dE \\
1038885\tablenotemark{h} & 2  1 & 21.36 & 21.86 & 22.10 & 0.77 & 0.76 & 22.80 & 22.87 & 12 59 38.62 & 27 56 58.64 &  6718 & 47 &  4.0 & & & 5.1 &  dE \\
122743 &  4 4  & 20.17 & 20.07 & 20.19 & 0.92 & 1.40 & 20.59 & 19.54 & 12 59 42.27 & 28 00 54.53 &  7934 & 77 & 15.0 & & & 40.7  &   E \\
195506 &  2  1 &       & 19.55 & 19.77 &      & 0.97 &       & 23.13 & 12 59 42.93 & 27 59 54.23 &  8497 & 83 &  7.9 & & &  11.3  &  dE \\
196718 &  2  1 & 21.05 & 20.82 & 21.10 & 0.65 & 0.90 & 23.09 & 23.56 & 12 59 43.94 & 27 58 01.95 &  7926 & 94 & 5.4 & & & 6.2  &   dE \\
250641 &  2  1 & 21.54 & 21.86 & 22.06 & 0.62 & 0.97 & 23.54 & 23.79 & 12 59 44.01 & 27 56 15.32 &  7316 & 85 &  3.2 & & & 5.4  &   dE \\
180052 &  2  x & 19.92 & 19.75 & 20.04 & 0.62 & 1.06 & 22.68 & 23.40 & 12 59 44.72 & 27 58 06.77 &  9870 & 74 & 4.2 & & & 10.4  &   dE \\
182103 &  3  2 & 20.54 & 20.56 & 20.80 & 0.74 & 1.08 & 22.18 & 22.99 & 12 59 51.83 & 27 57 26.25 &  5960 & 74 & 7.6 & & & 12.4  &   dE \\
4042826\tablenotemark{h} & 2 1 &  20.54 & 20.38 & 20.68 & 0.70  & 0.92 & 22.98  & 23.55 & 12 59 52.18 & 27 59 46.48 & 8641 & 116 & 2.2 & & & 5.0  &  dE \\
183846 &  4  3 & 20.07 & 20.02 & 20.19 & 0.64 & 0.86 & 21.86 & 22.41 & 12 59 59.97 & 27 59 17.88 &  9031 & 79 & 5.3 & & & 22.4 &   S0 \\
242439 &  x  3 & 20.36 & 20.20 & 20.38 & 0.76 & 1.05 & 21.73 & 21.64 & 13 00 10.38 & 27 56 17.08 &  7485 & 60 &  8.7 & & & 26.4  &   E \\
31001  &  3  1 & 21.83 & 21.80 & 22.06 & 0.53 & 0.69 & 23.36 & 23.57 & 13 00 11.96 & 28 04 03.01 &  7104 & 80 & 4.3 & & & 7.7  &   dE \\
101463 &  2  2 &       & 20.53 & 20.83 &      & 1.08 &       & 23.60 & 13 00 13.52 & 28 02 14.60 &  5327 & 65 &  7.4 & & & 8.1 &  dE \\
101935 &  3  2 & 21.57 & 21.59 & 21.91 & 0.71 & 0.88 & 23.30 & 23.77 & 13 00 14.21 & 28 00 26.38 &  8149 & 85 &  5.7 & & & 5.4 &   dE \\
101962 &  2  1 & 20.36 & 20.28 & 20.59 & 0.78 & 1.03 & 23.14 & 23.87 & 13 00 15.70 & 28 01 46.24 &  6985 & 92 &  3.8 & & & 5.2 &   dE \\
230614 &  1  1 & 19.55 & 19.54 & 19.78 & 0.66 & 0.90 & 22.61 & 23.22 & 13 00 16.36 & 27 55 22.11 &  4665 & 66 &  7.9 & & & 9.5  &   dE \\
230665 &  2  1 & 18.95 & 19.31 & 19.52 & 0.76 & 0.92 & 22.54 & 23.32 & 13 00 16.68 & 27 56 38.81 &  5366 & 78 &  6.5 & & &  9.9 &    dE\\
160495 &  3  2 & 21.10 & 21.17 & 21.42 & 0.67 & 0.89 & 23.27 & 23.73 & 13 00 18.26 & 27 58 17.51 &  4950 & 79 &  4.5 & & & 5.3 &   dE \\
230945 &  3  1 & 20.04 & 20.20 & 20.38 & 0.66 & 0.92 & 22.71 & 23.36 & 13 00 18.41 & 27 55 16.85 &  8155 & 74 & 5.1 & & &  11.0 &  dS0 \\
32410  &  2  1 & 21.82 & 21.39 & 21.69 & 0.61 & 1.00 & 23.34 & 23.58 & 13 00 19.32 & 28 04 22.27 &  7349 & 99 &  4.2 & & &  6.5 &  dE \\
160703 &  2  1 & 21.29 & 20.90 & 21.39 & 0.64 & 0.86 & 23.66 & 24.18 & 13 00 19.74 & 27 59 04.88 &  7035 & 135 &  2.9 & & & 5.7 &   dE \\
32493  &  2  1 & 20.44 & 20.31 & 20.53 & 0.70 & 0.95 & 22.62 & 23.13 & 13 00 20.28 & 28 04 53.15 &  8128 & 69 &  6.2 & & & 12.5 &   dE \\
90186  &  2  1 & 20.21 & 20.12 & 20.49 & 0.75 & 0.99 & 22.93 & 23.65 & 13 00 21.97 & 28 02 18.39 &  8051 & 63 & 6.7 & & & 4.7  &   dE \\
231908 &  2  1 & 19.49 & 19.28 & 19.53 & 0.74 & 1.00 & 22.56 & 23.14 & 13 00 22.95 & 27 55 15.30 &  5270 & 69 &  7.0 & & & 14.8  &  dE \\
161594 &  2  1 & 21.12 & 21.16 & 21.56 & 0.71 & 0.96 & 23.57 & 24.13 & 13 00 23.22 & 27 59 48.92 &  6321 & 75 &  3.4 & & & 4.6  &  dE \\
90335  &  2  1 & 19.29 & 19.34 & 19.53 & 0.77 & 0.95 & 21.58 & 22.88 & 13 00 23.47 & 28 03 02.00 &  7076 & 59 &  9.5 & & &28.5  &   dE \\
162112 &  3  1 & 19.75 & 20.82 & 19.89 & 0.71 & 0.92 & 21.97 & 23.08 & 13 00 24.85 & 27 59 21.90 &  9217 & 65 &   7.9 & & & 19.6&  dE \\
232063 &  3  1 & 19.98 & 19.93 & 20.15 & 0.77 & 1.04 & 20.15 & 22.98 & 13 00 25.05 & 27 56 38.00 &  6946 & 87 &  7.0 & & &13.0 &  dE \\
162484 &  3  1 & 20.34 & 20.28 & 20.54 & 0.71 & 1.03 & 22.76 & 23.16 & 13 00 26.76 & 27 59 53.57 &  7195 & 67 &  5.9 & & &8.6  &   dE \\
91543  &  4  3 & 19.03 & 18.83 & 18.98 & 0.99 & 1.38 & 20.07 & 20.05 & 13 00 27.34 & 28 00 33.26 &  6428 & 51 & 26.3 & & &43.4  &  E \\
163240 &  3  1 &       & 20.81 & 21.00 &      & 0.96 &       & 23.43 & 13 00 29.71 & 27 58 06.79 &  5220 & 71 & 5.9 & & & 8.6 &  dE \\
91844  &  3  1 & 20.81 & 20.60 & 20.87 & 0.71 & 0.95 & 23.00 & 23.48 & 13 00 30.02 & 28 01 35.08 &  6588 & 75 &  5.3 & & & 10.1 &  dE \\
91752  &  2  x & 20.53 & 19.90 & 20.34 & 0.68 & 0.92 & 22.71 & 23.81 & 13 00 30.93 & 28 03 12.76 &  7395 & 65 &  7.3 & & & 10.5  &   dE\\
92156  &  2  1 & 20.72 & 20.79 & 21.20 & 0.79 & 0.87 & 23.52 & 24.09 & 13 00 31.76 & 28 01 21.84 &  8123 & 86 &  3.7 & & & 7.0 &  dE  \\
150137 &  1  1 & 18.79 & 18.83 & 19.13 & 0.74 & 0.99 & 22.59 & 23.31 & 13 00 32.48 & 27 58 32.87 &  8435 & 106 &  3.5 & & & 10.4 &  dE \\
150530 &  4  1 & 20.32 & 20.21 & 20.43 & 0.73 & 1.00 & 22.52 & 23.10 & 13 00 34.29 & 27 58 17.63 &  7817 & 73 &  7.0 & & & 15.2 &  dE \\
221227 &  3  1 & 19.84 & 19.73 & 19.91 & 0.74 & 0.96 & 22.21 & 23.08 & 13 00 35.99 & 27 55 05.59 &  6946 & 59 & 9.9 & & & 15.7 &  dE \\
92663 &  x x   & 21.02\tablenotemark{e} & 23.00 & 23.13 & 0.46\tablenotemark{e} & -0.92 & 21.81  & 20.38 & 13 00 36.54 & 28 02 55.53 &    &       & &   7557 & 70 &  &   \\
  &     &       &       &      &         &       & &      &        &       &    &        & &   7506\tablenotemark{f} & 30 & &  \\
151072\tablenotemark{g} & 4 3 & 21.48 & 21.47 & 21.60 & 0.71 & 0.96 & 21.65 & 20.14 & 13 00 37.84 & 27 58 40.94 & 4906  & 74 &  6.7 & & & 18.9  &   E \\
 & &  &  & & &  & & & & &  4845  & 59 &  10.0 & & & 21.3  &  \\ 
80651  &  2  2 &       & 18.12 & 18.30 &       & 0.54 &      & 21.84 & 13 00 39.08 & 28 00 35.51 &  5932 & 76 &  5.4 & & & 59.1 &  Sa \\
151316 &  2  1 & 20.27 & 20.08 & 20.45 & 0.65 & 0.98 & 22.95 & 23.82 & 13 00 39.28 & 27 57 48.01 &  6599 & 73 &  3.4 & & & 5.7   &  dE \\
10162  &  1  1 & 20.45 & 20.45 & 20.75 & 0.68 & 0.93 & 23.15 & 23.81 & 13 00 39.37 & 28 04 11.37 &  6652 & 55 &  6.4 & & & 9.8  &  dE \\
10640  &  3  2 & 18.96 & 18.88 & 19.06 & 0.80 & 1.12 & 21.10 & 21.82 & 13 00 42.52 & 28 03 25.33 &  5817 & 64 &  13.8 & & & 32.6 &  E \\
151773 &  3  1 & 21.00 & 20.96 & 21.29 & 0.74 & 0.91 & 23.14 & 23.69 & 13 00 43.73 & 27 59 20.84 &  4840 & 80  & 7.2 & & & 7.8  &   dE \\
81390  &  2  1 & 18.68 & 18.86 & 19.10 & 0.75 & 1.02 & 22.19 & 23.13 & 13 00 44.10 & 28 02 15.43 &  8889 & 75 &  10.0 & & &  27.6  &  dE  \\
11874  &  3  1 & 20.98 & 21.35 & 21.75 & 0.69 & 0.93 & 23.39 & 24.01 & 13 00 45.92 & 28 03 35.46 &  7919 & 79  &  4.2 & & & 5.7  &  dE \\
150005 &  3  1 & 20.78 & 21.37 & 21.50 & 0.71 & 0.96 & 22.70 & 22.81 & 13 00 47.55 & 27 58 29.91 &  5820 & 61 & 8.2 & & &12.7  &  dE \\
12045  &  3  3 & 21.96 & 21.78 & 21.98 & 0.62 & 0.98 & 23.58 & 23.85 & 13 00 51.03 & 28 03 47.62 &  6678 & 72 & 5.6 & & & 7.4  &  dS0 \\

%\bf{Other} &  & & & & & & & & & & &  & & & \\
%92663 &  x x & 21.02\tablenotemark{d}  & 23.13 & 0.46\tablenotemark{d} & -0.93 & 21.81  & 13 00 36.54 & 28 02 55.53 &  9 &  203 & 184 &      &       &   & 7557 & 50 &  33.7 &  ?? \\
%  &     &       &       &      &         &             &        &       &    &      &     &       &         &   & 7506\tablenotemark{e} & 30 & 46.3 &  \\
%
\bf{Bckgrd} &  & & & & & & & & & & &  &  &    \\
132716 &  3  2 & 19.05 & 18.83 & 18.97 & 1.50 & 1.78 & 19.94 & 19.79 & 12 59 29.19 & 28 01 31.76 &  55023 & 101 & 14.1 &  54866 & 99 & 61.1 &  E\\
120551 &  3  2 & 22.01 & 21.68 & 22.05 & 0.79 & 0.83 & 23.40 & 23.68 & 12 59 34.78 & 28 02 01.36 &  49868 & 179 & 1.1 & 49484 & 72 & 7.3 & dE  \\
192871 &  4  3 & 19.92 & 19.59 & 19.74 & 1.13 & 1.68 & 20.49 & 19.63 & 12 59 37.13 & 27 58 46.93 &  115172 & 87 & 4.5 & 115228 & 74 & 34.8 & Sp \\
%195768 &  4  3 & 22.47 & 21.82 &  & 23.34 &   12 59 40.73 & 27 57 02.92 &  19 &  3663 & 908 & ? 0.856 &  4.9 &    &  \\
182393 &  3 x & 21.19 & 20.81 & 21.08 & 0.39 & 1.37 & 23.02 & 23.04 & 12 59 52.59 & 27 56 57.90 & &    &  &    208287 & 41 &  & \\
100600 &  4 4 & 22.17 & 21.63 & 21.82 & 0.89 & 2.06 & 23.40 & 22.80 & 13 00 09.01 & 28 02 44.84 & &    &  &    246765 & 70 & & Sp\\
101905 &  x x & 21.79 & 21.61 & 21.90 & 0.73 & 1.21 & 22.42 & 22.02 & 13 00 13.85 & 28 00 15.14 & &    &  &   138648 & 36 & & \\
230804 &  4 3 & 21.39 & 20.60 & 20.73 & 2.03 & 3.07 & 21.93 & 20.48 & 13 00 14.14 & 27 53 58.08 & 204700 & 63 & 10.1  & &  &  &  S0  \\
230475 &  3 3 & 19.25 & 19.21 & 19.41 & 0.79 & 1.02 & 21.58 & 22.57 & 13 00 16.28 & 27 56 30.48 &   49616 & 104 & 3.3 & 49669 & 135 & 15.1& Sp\\
161002 &  4  3 &      & 19.90 & 20.03 &      & 1.25 &       & 19.55 & 13 00 17.60 & 27 58 07.62 &  63908&  87 & 6.2 &  63817 & 90 & 44.0& E \\
 &   &  & &  &  &   & & &   &  & 63973 &  90 & 5.2 &  63953 & 76  & 40.6 & \\
160796 & 5  x & 23.41 & 22.91  & 23.20 &     & 1.33 & 24.12 & 23.19 & 13 00 18.66 & 27 58 17.10 &      &   &    &    245259 & 49 & &  \\
32265  &  x x & 21.57 & 21.30 & 21.54 & 1.01 & 1.22 & 22.71 & 22.61 & 13 00 17.67 & 28 03 34.53 & 82392 & 88 & 4.4 &  82282 & 71 & 8.9 & \\
160607 &  x x & 22.51 & 22.54 & 22.77 & 0.70 & 0.88 & 23.59 & 22.77 & 13 00 18.63 & 27 58 22.73 &    & &  &   111330 & 41 & &  \\
32751  &  4 4 & 20.75 & 20.20 & 20.37 & 1.04 & 2.63 & 22.03 & 21.62 & 13 00 20.62 & 28 04 15.69 & 192640 & 71 & 3.7 &  192765 & 43 & & Sp \\
230000  &  5 x &      &  &       &      &      &       & & 13 00 22.84 & 27 55 15.52 &   & &  &  210243 & 39 & &  \\
232587 &  4 3 & 22.19 & 21.67 & 21.86 & 0.35 & 1.15 & 22.87 & 22.37 & 13 00 27.17 & 27 54 53.29 &  &   &  &     227441 & 41 &  & Irr \\
220952 &  x x & 21.87 & 21.14 & 21.26 & 0.91 & 2.18 & 22.12 & 20.28 & 13 00 34.25 & 27 54 47.15 & 259047 & 83 & 2.8 &  259248 & 87  &  & \\
22382  &  4 3 & 21.19 & 20.41 & 20.54 & 1.62 & 3.00 & 21.82 & 20.69 & 13 00 35.74 & 28 03 19.44 & 192345 & 91 &  3.0&  & &    &   E \\ 
221307 &  4 3 & 20.78 & 20.51 & 20.67 & 0.47 & 1.38 & 21.17 & 20.48 & 13 00 36.22 & 27 55 34.60 &      &  &    & 192243 & 34 &  & Irr \\
80801  &  x x &       & 21.78 & 21.96 &      & 2.48 &       & 21.53 & 13 00 39.68 & 28 00 41.69 &  138122 & 56 &  4.7 & & & &   \\
80686  &  x x & 22.53 & 21.10 & 22.25 & 0.64 & 1.54 & 22.33 & 20.62 & 13 00 39.74 & 28 01 11.89 &   &     &   &   176421 & 32 &  & \\
80928  &  x x & 22.01 & 21.62 & 21.81 & 0.69 & 1.35 & 22.72 & 21.87 & 13 00 42.36 & 28 02 41.35 &   &  &  &   228058 & 38 & &  \\
%151841 &  3  x & 22.36 & 22.44 & 0.67 & 23.46 &   13 00 43.79 & 27 59 00.02 &  15 &  1339 & 1224 & ? 0.165 & 1.0& ? 0.164  &  \\
10863  &  x x & 21.64 & 21.45 & 21.60 & 0.30 & 0.89 & 22.13 & 21.57 & 13 00 45.15 & 28 05 20.63 &   &  &  &   192212 & 31 & & \\

\bf{Uncertain} & &  & & & & & & & & & &  & &    \\
133528 &  x x & 21.81 & 21.44 & 21.60 & 0.71 & 2.25 & 22.18 & 20.85 & 12 59 31.82 & 28 00 32.81 &  154006 & 91 &  3.1 & & &  &   \\
181737 &  x x & 22.11 & 21.91 & 22.01 & 0.35 & 1.01 & 22.24 & 20.43 & 12 59 53.19 & 28 00 10.57 &  &  & &  295925 & 32 & & \\
100940 &  4 3 & 21.03 & 20.94 & 21.06 & 0.98 & 2.14 & 21.63 & 20.47 & 13 00 09.79 & 28 01 37.19 & 125875 & 63 & 5.1 & & &    &  E \\
%230751 &  2  1 & 21.61 & 23.67 & 0.70 & 23.76 &   13 00 16.24 & 27 54 44.07 &  23 &  3070 & 3120 & 0.020 &  2.8 &      &  \\
32910  & 1  1 & 20.88 & 20.85 & 21.23 & 0.67 & 1.00 & 23.63 & 24.02 & 13 00 21.35 & 28 03 26.87 &  7870 & 91 &  3.3 & & & 3.7 &   dE \\
%32799  &  3  3 & 22.03 & 22.61 & 0.77 & 23.09 &   13 00 21.79 & 28 04 41.48 &   3 &  2035 &  566 &  0.029 &  5.4 &  0.470 &   \\
90680  & 2  2 & 21.18 & 21.27 & 21.58 & 0.63 & 0.96 & 23.35 & 23.90 & 13 00 22.85 & 28 00 56.61 &  9272 & 90 & 2.9 & & & 5.8  &  dE \\
22069  & 3  3 & 22.38 & 22.02 & 22.25 & 0.84 & 1.48 & 23.24 & 23.03 & 13 00 34.77 & 28 04 32.88 &        &    &  &   117282 & 76  & & Sp \\
80942  &  x x & 21.89 & 21.26 & 21.43 & 0.24 & 0.85 & 22.49 & 22.13 & 13 00 39.94 & 28 00 24.08 &    & &  &  276525 & 31 & &  \\
10217  &  x x & 22.11 & 21.49 & 21.66 & 0.43 & 1.18 & 22.51 & 21.48 & 13 00 40.36 & 28 04 54.48 &    & &   &  310392 & 31 & & \\

\bf{Failed} & &  & & & & & & & & & & & &    \\
\bf{(Poss. mem)} & &  & & & & & & & & & & & &    \\
190565 & 2 1 & 22.00 & 23.44 & 23.64 & 0.73 & 0.69 & 23.77 & 23.09 & 12 59 32.75 &  27 59 26.34 & & & & & & &  dE \\
182496 & 3 3 &       & 22.35 & 22.66 &      & 0.81 &       & 23.51 & 12 59 53.74 &  27 58 21.64 & & & & & & &  Sp  \\
100420 & 1 1 & 22.20 & 22.84 & 23.53 &      & 0.82 & 24.09 & 24.17 & 13 00 06.89 &  28 01 55.89 & && & & & & dE \\
242067 & 2 x & 21.77 & 21.64 & 22.17 & 0.71 & 0.54 & 23.86 & 24.28 & 13 00 07.95 &  27 56 35.60 && & & & & &  dE\\
30000  & 1 x & 22.12 & 21.35 & 21.61 & 0.73 & 0.88 & 23.40 & 23.92 & 13 00 11.78 &  28 05 05.00 & &  &  &   & & &  dE\\
230751 & 2 1 & 21.61 & 23.51 & 23.67 & 0.70 & 0.85 & 23.76 & 22.68 & 13 00 16.24 &  27 54 44.07 & & &   &   & & &   dE\\
160792 & 3 2 & 22.17 & 22.19 & 22.62 & 0.69 & 0.84 & 23.83 & 24.12 & 13 00 18.93 &  27 58 00.55 & &  &  & & & &  dE \\
102885 & 1 1 & 21.86 & 21.94 & 22.49 & 0.79 & 0.95 & 23.89 & 24.44 & 13 00 20.27 &  28 02 02.15 & & & & & & &   dE \\
32799  & 3 3 & 22.03 & 22.37 & 22.62 & 0.77 & 1.32 & 23.09 & 22.73 & 13 00 21.79 &  28 04 41.48 & & &   & & &  &  cE \\
20411  & 3 x & 22.45 & 21.91 & 22.13 & 0.31 & 1.25 & 23.70 & 23.43 & 13 00 24.62 &  28 03 42.86 & &  &  & & & &  \\
160003 & 2 x & 21.85 & 21.62 & 22.19 & 0.67 & 0.84 & 23.93 & 24.37 & 13 00 33.01 &  27 59 44.05 & &  &  &   & & &  \\

\bf{Failed} & &  & & & & & & & & & & & &   \\
\bf{(Poss. bkg)} & &  & & & & & & & & & & & &   \\
195768 & 4  3& 22.47 & 21.68 & 21.82 &      & 3.43 & 23.34 & 21.50 & 12 59 40.73 & 27 57 02.92 & & &  &   & &   & E \\
161665 & 4 3 & 22.17 & 21.29 & 21.51 & 0.68 & 1.87 & 23.22 & 22.78 & 13 00 20.50 & 27 57 00.08 & & & & & & &  Sp \\
162881 & 4 3 & 22.13 & 21.76 & 21.93 &      & 3.47 & 23.48 & 22.41 & 13 00 28.55 & 27 58 59.20 & & & & & & & E \\
151841 & 3 x & 22.36 & 22.33 & 22.44 & 0.67 & 0.56 & 23.46 & 23.89 & 13 00 43.79 & 27 59 00.02 & &  & &  & &   &  \\

\enddata

\tablecomments{Reddening is not taken into account here, but typical E$_{(B-V)} \sim 0.01$.}
\tablenotetext{a}{Membership probability classes of NT and HF.  An
x indicates no membership probability was assigned.  In most cases these objects were 
assumed to be background. A 5 denotes that the galaxy was an extra object that landed
in the slit.}
\tablenotetext{b}{$BVR$ photometry from catalog of \citet{adami}}
\tablenotetext{c}{Errors include contributions from the cross correlation measurement
uncertainty and from the uncertainties in the sky line wavelength calibration shift.}
\tablenotetext{d}{Signal-to-noise ratio per Angstrom around 5000\AA}
\tablenotetext{e}{Photometry is for two blended components including the UCD
target and another source centered 0.9 arcsec west of it.}
\tablenotetext{f}{Redshift measurement from red-side data}
\tablenotetext{g}{Object was observed in two masks.}
\tablenotetext{h}{Object was observed in a separate run in 2009}
\end{deluxetable}

%\end{document}

%-------------------------------------------------
% FIGURES
%--------------------------------------------------

\clearpage

\clearpage

\end{document}